\documentclass[preprint,12pt,showpacs,aps,amsmath,amssymb, 
nofootinbib,superscriptaddress,hyperref,showkeys]{revtex4} 

\usepackage[utf8]{inputenc}

\usepackage{epsfig} 
\usepackage{wasysym} 
\usepackage{mathrsfs} 
\usepackage{graphicx} 
\usepackage{amsfonts} 
\usepackage{amsbsy} 
\usepackage{amscd} 
\usepackage{pstricks} 
\usepackage{multirow}
\usepackage{slashed} 
\usepackage{tikz}
\usetikzlibrary{decorations.pathmorphing,decorations.markings}
 \usepackage{amssymb}
\usepackage{mathtools}
 \usepackage{float}

\def\issue(#1,#2,#3){{\bf #1} (#2) #3} 

\def\APP(#1,#2,#3){Acta Phys.\ Polon.\ \issue(#1,#2,#3)}
\def\ARNPS(#1,#2,#3){Ann.\ Rev.\ Nucl.\ Part.\ Sci.\ \issue(#1,#2,#3)}
\def\CPC(#1,#2,#3){Comp.\ Phys.\ Comm.\ \issue(#1,#2,#3)}
\def\CIP(#1,#2,#3){Comput.\ Phys.\ \issue(#1,#2,#3)}
\def\EPJC(#1,#2,#3){Eur.\ Phys.\ J.\ C\ \issue(#1,#2,#3)}
\def\EPJD(#1,#2,#3){Eur.\ Phys.\ J. Direct\ C\ \issue(#1,#2,#3)}
\def\IEEETNS(#1,#2,#3){IEEE Trans.\ Nucl.\ Sci.\ \issue(#1,#2,#3)}
\def\IJMP(#1,#2,#3){Int.\ J.\ Mod.\ Phys. \issue(#1,#2,#3)}
\def\JHEP(#1,#2,#3){JHEP\ \issue(#1,#2,#3)}
\def\JCAP(#1,#2,#3){JCAP\ \issue(#1,#2,#3)}
\def\JPG(#1,#2,#3){J.\ Phys.\ G \issue(#1,#2,#3)}
\def\MPL(#1,#2,#3){Mod.\ Phys.\ Lett.\ \issue(#1,#2,#3)}
\def\NP(#1,#2,#3){Nucl.\ Phys.\ \issue(#1,#2,#3)}
\def\NIM(#1,#2,#3){Nucl.\ Instrum.\ Meth.\ \issue(#1,#2,#3)}
\def\PL(#1,#2,#3){Phys.\ Lett.\ \issue(#1,#2,#3)}
\def\PRD(#1,#2,#3){Phys.\ Rev.\ D \issue(#1,#2,#3)}
\def\PRL(#1,#2,#3){Phys.\ Rev.\ Lett.\ \issue(#1,#2,#3)}
\def\PTP(#1,#2,#3){Progs.\ Theo.\ Phys. \ \issue(#1,#2,#3)}
\def\RMP(#1,#2,#3){Rev.\ Mod.\ Phys.\ \issue(#1,#2,#3)}
\def\SJNP(#1,#2,#3){Sov.\ J. Nucl.\ Phys.\ \issue(#1,#2,#3)}
\def\PR(#1,#2,#3){Phys.\ Rep.\ \issue(#1,#2,#3)}

\allowdisplaybreaks

\begin{document}

\author{Selvaganapathy J}
\email{jselva@prl.res.in, jselva1729@gmail.com}
\affiliation{Theoretical Physics Group, Physical Research Laboratory, Ahmedabad 380 009, India.}

\title{Pure Natural Quintessential Inflation and Dark Energy}


\begin{abstract}
We propose the pure natural quintessential inflation model which is motivated by Witten's conjecture, where the axion couples to pure Yang-Mills $SU(N)$ gauge field at large N limit. This modifies the standard cosine potential which is presented in the natural inflation, making it compatible with current CMB data. Our model gives a successful inflation as well as acceleration at the late times  by quintessence inflaton field($\phi$). Here the inflaton field is responsible for inflation, after that the field enters into peculiar type of reheating and then they act as dynamical dark energy field which follows the same inflation potential and same model parameters. The dynamical field slowly rolls until the Hubble drops to mass of the quintessence field and it reaches the current dark energy field value. Here the dark energy scale is field dependent. Our quintessence model follows thawing frozen approach therefore the frozen quintessence field evolves with respect to cosmic time from initial field value($\phi_{i}$) to present non-zero minimum field value($\phi_{0}$). The obtained field value turned into ultra-light and  it satisfies the present dark energy density which is $V(\phi_{0}) \approx \Lambda_{\text{DE}}^{4} = (2.3 \times 10^{-3}\text{eV})^{4}$.
\end{abstract}


\maketitle

\section{Introduction}
\label{sec:intro}
\noindent It is believed that the universe started with initial conditions seeded by cosmic inflation, which grew up the observable early universe at very short time\cite{guth,linde}. The inflationary approach remarkably alleviates the infamous horizon and flatness problem of hot Big-Bang universe at early time. At the end of inflation, the energy density of the inflaton transform to other degrees of freedom which are standard model particles, dark matter and dark energy. Here the baryonic matter and dark matter would be the responsible for the galaxy formation at low redshift era. Meanwhile, the amount of their energy densities are highly constrained by Big-bang nucleosynthesis(BBN) as well as initial conditions of the cosmic inflation when they produced at early time(high redshift). Though the dark matter drives the growth of the structure formation faster, dark energy slows down in it due to their property of negative pressure at late time nearly low redshift;
larger the acceleration, higher the suppression of growth of structure \cite{strucform,strucform1,strucform2,strucform3}. Thus the radiation dominated and matter dominated phases of universe sandwiched by early and late time cosmic acceleration, both acceleration drives the universe by negative pressure through violating the strong energy condition(SEC). Such kind of late time cosmic acceleration confirmed the existence of dark energy by type Ia supernovae distance measurement as a cosmic candle\cite{planck1,planck2}. 

\noindent Importantly the measurement of the baryon acoustic oscillation(BAO) is the standard ruler which exhibit the distribution(not random) of the energy density fluctuation of the matter at a very large scale\cite{bao1,bao2}. Even though the matter distributed in a non-linear way as clumps at small scale($<10$ Mpc) which is not surrounded by empty space instead such void space occupied by dark energy homogeneously and isotropically. Two different kinds of candidate explain the nature of dark energy with negative pressure which are cosmological constant and quintessence\cite{ratra,watteric,steinhardt1,dde1,watterich1};
the vacuum energy density is constant with fixed equation of state(EOS) $\omega=-1$ throughout universe evolution which is known as cosmological constant, instead the energy density dynamically vary with $\omega$(close to $-1$) as well as cosmic scale factor called as quintessence. Apart from equation of state $\omega$ of the dark energy, the sound speed $C_{s}$ is also important to distinguish whether the dark energy belongs to cosmological constant or quintessence. The sound speed affects the quintessence energy density distribution as well as power spectrum of cosmic micro wave background(CMB), if it differ from unity\cite{soundspeed1}. But most of the quintessence model takes $C_{s}=1$ which means sound speed equal to the speed of light, thereby the effect of this property is confined to very large scale, this can be manifested in the large-angle multipoles of the CMB anisotropies.
Such probes of large angular multipoles(low $l$) in the CMB anisotropies, the equation of state measurement of supernovae type-Ia, large scale structure(LSS) and CMB which results $\omega\neq-1$ with error, could be favored to dynamical dark energy model.


\noindent In this perspective, the early and late time cosmic acceleration can be unify through quintessence field model. There are few models proposed in this direction which satisfies slow-roll inflation data as well as provides the quintessence nature of dark energy. In this regard, most importantly Starobinsky model\cite{starobinsky,helical}, Natural inflation\cite{ni} and Axion monodromy model\cite{monodromy1,monodromy2} satisfy the recent inflationary data(Planck2018) within the $3\sigma$.\\
\noindent During the inflation period the wavelength of the comoving scale are exponentially streched compared to Hubble radius. According to CMB anisotropy measurement and present large scale structure, the wavelength of the comoving scale expanded about 60 times of the Hubble radius at the end of the inflation. Thus the observed upper limit of the tensor-scalar ratio in the Planck 2018 result which exhibits that the cosmological field space resides on the super-Planckian in accord with Lyth bound.

\noindent The gauge field theory of standard model(SM) of particle physics and the quantum theory of gravity which is predictive effective field theory(EFT) are valid below the planck scale. Therefore EFT of gravity and SM breaksdown at super-Planckian scale and also expected to breaks at some UV scale. However gravity is stronger at Planck scale as well as super-Planckian. Therefore there is no precise quantum theory of gravity to understand the cosmological structure. But we can classify the inflationary model by weak gravity conjecture, whether they are UV completed or not and whether they are valid at sub-Planckian or super-Planckian. We do not have any proof of such conjecture but which is inpired by blackhole physics. The scalar weak gravity conjecture states that the scalar interaction strength is stronger than the gravitational interaction strength at below the Planck scale. Here one can relate the UV and IR scalar physics with gravitational strength of the scalar in minimally coupled gravitational theory. 

\noindent In this paper, we have explore the aspects of natural inflation with large-N limit pure Yang-Mills theory. In section.\ref{sec2}, we present the analytical expression for axionic potential which is based on Witten's conjecture, then we perform the numerical analysis of our model by satisfying Planck 2018 results. Further we have discussed about sub-planckian as well as super-Planckian field physics through Lyth bound. Section.\ref{sec3} includes the weak gravity conjecture. Here we get the constraint for tensor-scalar ratio from swampland conjecture though it conflicts with slow-roll condition. Then we have analyzed the strong scalar weak gravity conjecture in terms of slow-roll parameter up to fourth order which provides the criteria whether the inflation model valid up to super-Planckian field space or not. In section.\ref{sec4}, we have explained about the solution of dynamical dark energy parameters, depending on their properties we conclude that our model resides in the thawing frozen model. Here we have explained the nature of thawing dynamical dark energy by using our inflationary model results which are obtained from section.\ref{sec2}. Finally we have concluded our results in section.\ref{conclusion}.
\section{Large N dynamics potential and inflation}\label{sec2}
\noindent The large-N expansion of $SU(N)$ Yang-Mills theory was first studied by t'Hooft \cite{hooft} which characterized by local or global symmetry and their internal degrees of freedom which is related to a parameter N. In the case of QCD, the large-N limit doesn't make them solvable analytically in the four dimensional spacetime. According to t'Hooft mechanism, \textit{the $SU(N)$ Yang-Mills (YM) theory with the infinite number of color ($N\rightarrow\infty$), the t'Hooft coupling $\lambda\,=\,g^{2}\,N$ is fixed when the gauge coupling $g$ vanishes}. Thus the pure Yang-Mills interaction can be written in terms of expansions of $1/N$. But the shape of the instanton potential which generated at strong CP QCD vacuum cannot be same, meaning that the Cosine form of instanton potential is not valid in the large-N limit\cite{witten}.
%
Thus the Witten's conjecture states that the potential of the $SU(N)$ vacuum energy has to be i) multi-valued due to the existence of several meta-stable vacua, ii) the quadratic term (mass term) should be independent of number of color $N$, iii) the potential has to be continuous, smooth, periodic and CP invariant(before symmetry breaking). 
The $SU(N)$ Lagrangian in the large-N limit is
\begin{equation}
 \mathcal{L}= -\frac{1}{4}\left( \frac{N}{\lambda}\right) F_{\mu \, \nu} F^{\mu \, \nu} + \frac{\theta}{32 \pi^{2}} \, F_{\mu \, \nu} \widetilde{F}^{\mu \, \nu} \label{lagra1}
\end{equation}
Where $\widetilde{F}^{\mu \, \nu}= (1/2) \epsilon^{\rho \, \sigma \, \mu \, \nu}F_{\rho \, \sigma}$.  Here $\theta(=N \, \psi)$ is the angular variable. When
$ N \rightarrow \infty$, the $\psi$ kept fixed at $\theta =0$. The vacuum energy is minimum at  $\theta \rightarrow 0$ and the Euclidean path integrals gets 
real and positive. Importantly, when $\theta \neq 0$, the vacuum energy is minimized by maximizing the Euclidean space path maximal as a result the vacuum energy $E(\theta)$ computed by expanding the exponential $\exp(-S_{E})$. Where $S_{E}$ is the Euclidean action in the $\mathbb{R}^{4}$.
\begin{equation}
 \exp[-E(\theta)\, \text{Vol}(\mathbb{R}^{4})]=\int \, [D A_{\mu}]\,\exp(-S_{E})
\end{equation}
Thus the conjectured vacuum energy can be written as
\begin{eqnarray}
 E(\theta) & = & C \, \underset{k}{\min} (\theta+2 \pi k)^{2} + \mathcal{O}(1/N) \nonumber \\
 & \Rightarrow & N^{2}\, \underset{k}{\min}\, h\left(\frac{\theta+2 \pi k}{N}\right) \label{Witten}
\end{eqnarray}
The true vacuum of the multi branched vacuum energy is determined by certain branch value of $k$. Now we can able to solve the strong CP problem by introducing the Peccei-Quinn(PQ) global symmetry \cite{pecci}. According to PQ mechanism, the angular variable $\theta$ behave like a shift parameter to the axion field known as $a(x)$-pseudo Nambu-Goldstone boson with the periodicity of $2\, \pi\, f_{a}$. Here $f_{a}$ is the axion decay constant. Consequently, the instanton generates the $SU(N)$YM potential for axion-like inflaton field $a(x)$ by replacing $\theta = a(x)/f_{a}$ in the equation.(\ref{lagra1}), the potential for single branch can be written as follows \cite{witten}
\begin{equation}
 V(a(x))= N^{2} \Lambda^{4}\, g( \xi ) \, \, \quad \text{Where} \, \, \xi = \frac{\lambda\, a(x)}{8\, \pi^{2}\, N \,f_{a}} \label{Witten1}
\end{equation}
But the functional form of the inflaton potential in equation.(\ref{Witten1}) is not periodic under $a(x) \rightarrow a(x)+ 2\, \pi \, f_{a}$, thereby the large-N limit field parameter $\xi$ allows to construct a axionic inflaton model. Aside from lattice gauge theory, the string theory is the good candidate for constructing such axionic inflation model. Recently, the multi-valued branch axionic model was
proposed in ref.\cite{pni1,pni2,pni3} known as pure natural inflation inspired by axion monodromy \cite{monodromy1,monodromy2}. On the other hand, one can think Quintessence models \cite{steinhardt1,carrol} which explains inflation as well as dark energy i.e early as well as late time acceleration with the same potential and same value of model parameters. In this paper, we infer such kind of pure natural inflation model which is inspired by Quintessence in the context of large-N dynamics of $SU(N)$YM. Accommodating the Witten's conjecture(\ref{Witten}) the axionic potential can be written as follows 
\begin{equation}
 V(a(x)) = M^{4}\, \left[1-\exp \left[- \left( \frac{a(x)}{F_{a}}\right)^{2}\right] \right] \label{pni}
\end{equation}
Here $M=\sqrt{N} \Lambda$ and $F_{a}=N f_{a}$. Here the t'Hooft coupling $\lambda$ taken to be $8\pi^{2}$ and then the mass of the inflaton at large-N limit
obtained as $m_{a}= \sqrt{2} \Lambda^{2}/ f_{a}$. Here $\Lambda$ is the scale where the global symmetry has broken explicitly at lower energy scale 
compared to the spontaneous symmetry breaking scale $f_{a}$. 
\subsection{Planck result and Lyth bound}
\noindent The required curvature of the extremely flat potential for slow roll inflation is constrained by Lyth bound \cite{lyth1,lyth2,lyth3} which is 
\begin{equation}
 \frac{\Delta a(x)}{M_{Pl}} \gtrsim N_{e} \sqrt{\frac{r}{8}}. \label{lyth}
\end{equation}
Here $\Delta a(x)$ is the axion field traversing distance in the field space and $N_{e}$ is the required amount of e-folds for slow roll inflation and $r$ is the tensor-scalar ratio. The slow-roll parameters can be written as follows
\begin{equation}
 \epsilon_{V}=\frac{1}{2} \left(\frac{V'}{V}\right)^{2}; \quad \eta_{V}=\left( \frac{V''}{V}\right); \label{srlc1}
\end{equation}
Here prime represents the derivative with respect to field. The spectral tilt($n_{s}$), tensor-scalar ration($r$) and number of e-fold defined as follows 
\begin{equation}
 n_{s}= 1- 6 \epsilon_{V}+ 2\eta_{V}; \quad  r = 16 \epsilon_{V}; \quad N_{e}=\int^{a(x)_{\text{ini}}}_{a(x)_{\text{end}}} \frac{d a(x)}{ \sqrt{2 \epsilon_{V}}}; \nonumber
\end{equation}
 The non-zero $r$ emphasis that the possible existence of B-modes in CMB. The current Planck 2018 data \cite{planck1,planck2} has observed the upper bound of tensor-scalar ratio that is $ r_{0.002} < 0.064$ at $95 \% C.L$. 
\begin{figure}[h]
\centering
\includegraphics[width=0.75\textwidth,height=0.45\textwidth]{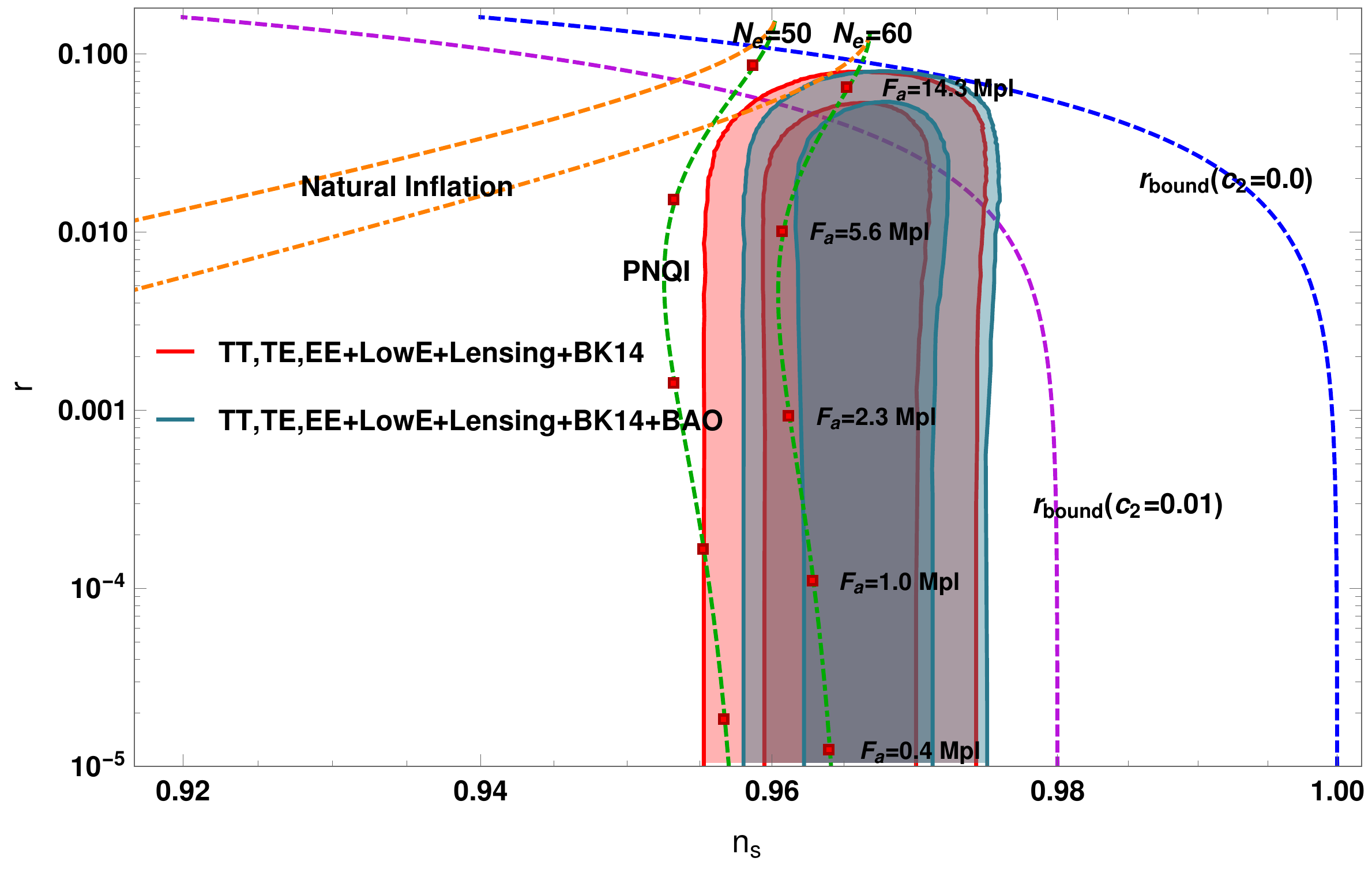} 
  \caption{The Pure natural quintessence inflation (PNQI) is consistent with Planck 2018 results with a wide range of tensor-scalar ratio for $60$ e-folds. Here $F_{a}=N\,f_{a}$. The natural inflation may be disfavour near the future for low $r$ value.}
   \label{fig:planck2018}
\end{figure}
This upper bound breaks the effective field theory when the Lyth bound at 60 e-folds which is $ \Delta a(x) \gtrsim 5.4 M_{Pl}$ thereby super-Planckian. In the case of natural inflation, to satisfy $r\approx 10^{-2}$, we require the axion decay constant $f_{a}$ to be $10 M_{Pl}$ as a result the single field natural inflation resides in the super-Planckian instead of sub-plankian. The most favored Starobinsky model satisfies spectral tilt($n_{s}$) $0.964$ nearly $r \sim 0.003$ for $55$e-folds. 
 \begin{table}[b]
\centering
 \begin{tabular}{|c|c|c|}
\hline
 Tensor-scalar ratio ($r$) & $f_{a}$(Unit of $M_{Pl}$) & Spectral tilt ($n_{s}$)   \\ 
 for e-fold $N_{e}=(50, 60)$ &  & for e-fold$N_{e}=(50, 60) $\\ 
 \hline
$(8.47\times 10^{-2}\, , \, 6.36\times10^{-2})$ & $14.3/N$ & $(0.9587\, , \, 0.9653)$  \\
$(1.45\times 10^{-2}\, , \, 9.91\times10^{-3})$ & $5.6/N$ & $(0.9533\, , \, 0.9607)$  \\
$(1.39\times 10^{-3}\, , \, 9.13\times10^{-4})$ & $2.3/N$ & $(0.9532\, , \, 0.9612)$   \\
$(1.63\times 10^{-4}\, , \, 1.08\times10^{-4})$ & $1.0/N$ & $(0.9553\, , \, 0.9628)$   \\ 
$(1.81\times 10^{-5}\, , \, 1.22\times10^{-5})$ & $0.4/N$ & $(0.9567\, , \, 0.9640)$   \\ 
\hline
\end{tabular}
\caption{Our Pure natural quintessence inflation (PNQI) at large-N limit satisfies wide range of tensor-scalar ratio for $50$ and $60$ e-folds. The sub-Planckian scale $f_{a}$ achieved by choosing $N \geq 15$, which is number of color presented in the $SU(N)$YM theory when present upper bound on tensor-scalar ratio $r<0.064$.} \label{tab:qpni}
\end{table}
One can argue that if the upper bound on $r$ lie beyond $10^{-2}$ or non-observation of B-modes in CMB or pointedly $r$ lie in the Big Bang Observer (BBO) near future \cite{BBO1,BBO2,BBO3,BBO4} which is $r \sim 10^{-3}-10^{-4}$ then one cannot have super-Planckian in the slow roll inflation. Once we get into sub-Planckian, we realize that the flatness and stability of the small field excursion inflation potential spoiled by radiative corrections in the low energy EFT. 
But the single field axionic inflation has global shift symmetry which preserves the flatness and stability of the potential at small field excursions \cite{nisymmetry1,nisymmetry2}.
In our model, the spontaneous and explicit broken symmetry scales of axion are well controlled by global symmetry and they are resides on sub-planckian. The figure.(\ref{fig:planck2018}) shows that our PNQI model satisfies the Planck2018 result within $3\sigma$ similar to the $\alpha$-attractor and the Starobinsky($\alpha=1$) model. The natural inflation doesn't satisfy the tensor-scalar ratio when it approaches lower than $10^{-2}$. The PNQI model valid for any range of lower value of $r$ which is shown in table.(\ref{tab:qpni}) and it will not cause any stability problem. To conclude here, the sub-planckian field theory where the symmetry breaking scale less than $M_{\text{Pl}}$ would be better handle than super-planckian to understand cosmology through field theory(EFT) point of view. In such case large-N limit natural inflation(our PNQI) model satisfy wide range of Lyth bound.     

\section{Weak gravity conjecture}
\label{sec3}
\subsection{Swampland conjecture}
\noindent The recent construction of conjectures \cite{swampland1,swampland2,swampland3} on string landscape and string swampland namely
\textit{de-Sitter swampland conjectures}, categorize all possible low energy effective QFT into UV-completed and non-UV completed in the context of quantum gravity. 
The necessary condition for the existence of UV completion field theory which as conjectured as follows\cite{swampland4}
\begin{enumerate}
 \item Distance conjecture:
 The range of scalar field traversed in the field space restricted as
 \begin{equation}
   \frac{| \Delta \phi  |} { M_{Pl}} \lesssim c_{0}
 \end{equation}
 \item Refined de-Sitter conjecture: Any scalar field with potential $V(\phi)$ in the low energy effective theory of consistent quantum gravity 
 must satisfy either 
 \begin{equation}
 \frac{| \nabla V  |} { V} \geq \frac{c_{1}}{M_{Pl}}  \label{rsc1}
 \end{equation}
or
 \begin{equation}
  min(\nabla_{i} \nabla_{j}V)\leq -\frac{c_{2}}{M_{Pl}^{2}}. V \label{rsc2}
 \end{equation}
\end{enumerate}
Where $c_{0}, c_{1}$ and $c_{2}$ are positive universal and $ \mathcal{O}(1) $ parameters. The parameter $c_{1}$ depends on the details of string flux compactification and it should be greater than $\sqrt{2}$. Here the de-Sitter conjecture doesn't allow (meta-)stable vacua with positive energy density and which is conflict with slow roll inflationary scenario \cite{swampland5,swampland6}. Thus the parametric constraints does not require to be $\mathcal{O}(1)$ rather it can be used to constrain the inflation paradigm. The distance conjecture do not have significant tension with the present observations(Lyth bound) but refined swampland conjecture(RSC) has non-trivial implications on potential dominated ($\dot{\phi}^{2}\ll V(\phi)$) slow roll inflation. According to RSC, the slow roll parameters defined customarily as
\begin{equation}
 \epsilon_{V} \geq \frac{c_{1}^{2}}{2}; \quad \eta_{V} \leq -c_{2} 
\end{equation}
In contrast to RSC, the slow roll inflation where $\epsilon_{V} \ll 1$ and $\eta_{V}\ll 1$ put restriction on the universal parameters $c_{1}$ and $c_{2}$ 
which cannot have $ \mathcal{O}(1) $ parameter due to current observations\cite{planck1,planck2} and their bounds are $c_{1}<0.1$ and $c_{2}<0.01$. Thereby the stability of de Sitter vacuum is questionable not only from observation, also from tensor perturbations \cite{tensfluc}, scalar entropic fluctuations \cite{scalarfluc1,scalarfluc2,scalarfluc3} and IR instability \cite{irinstable1,irinstable2}. The extensive study about de-sitter vacua with cosmological constant and neutrino mass made in \cite{ibanez}. Other hands, RSC equation.(\ref{rsc2}) with non-$\mathcal{O}(1)$ parameter $c_{1},\, c_{2}$ leads to the 
bound on tensor-scalar ratio \cite{phenorsc}
\begin{equation}
 r_{\text{bound}} \leq \frac{8}{3} (1-2\,c_{2}-n_{s}).
\end{equation}
The upper bound on $r$ shown in figure.(\ref{fig:planck2018}) for the values of $c_{2}=0 \, \text{and}\, 0.01$. In our model(equation.(\ref{pni})), the RSC equation.(\ref{rsc1}) can be written as axion field $a(x)$ which sets the upper limit of the axion decay constant, when $V''= 0$ at $a(x)/N f_{a} = 1/\sqrt{2}$ and the upper bound can be written as 
\begin{equation}
 f_{a} < \frac{2.18}{N . c_{1}} M_{\text{Pl}}.
\end{equation}
It is obvious that $N . c_{1}$ could be achieved greater than $2.18$ by demanding higher values of $N$ which keeps $f_{a} < M_{\text{Pl}}$ which is natural inflationary behaviour of the axion decay constant. Thus non-existence of de-sitter vacua or existence of quasi de-sitter vacua put a strong restriction on non-$\mathcal{O}(1)$ parameters by violating RSC which helps us to get the upper limit of the $f_{a}$ in our model with slow-roll inflation condition.
\subsection{Scalar weak gravity conjecture}
\begin{figure}[h]
\centering
\includegraphics[width=0.35\textwidth,height=0.28\textwidth]{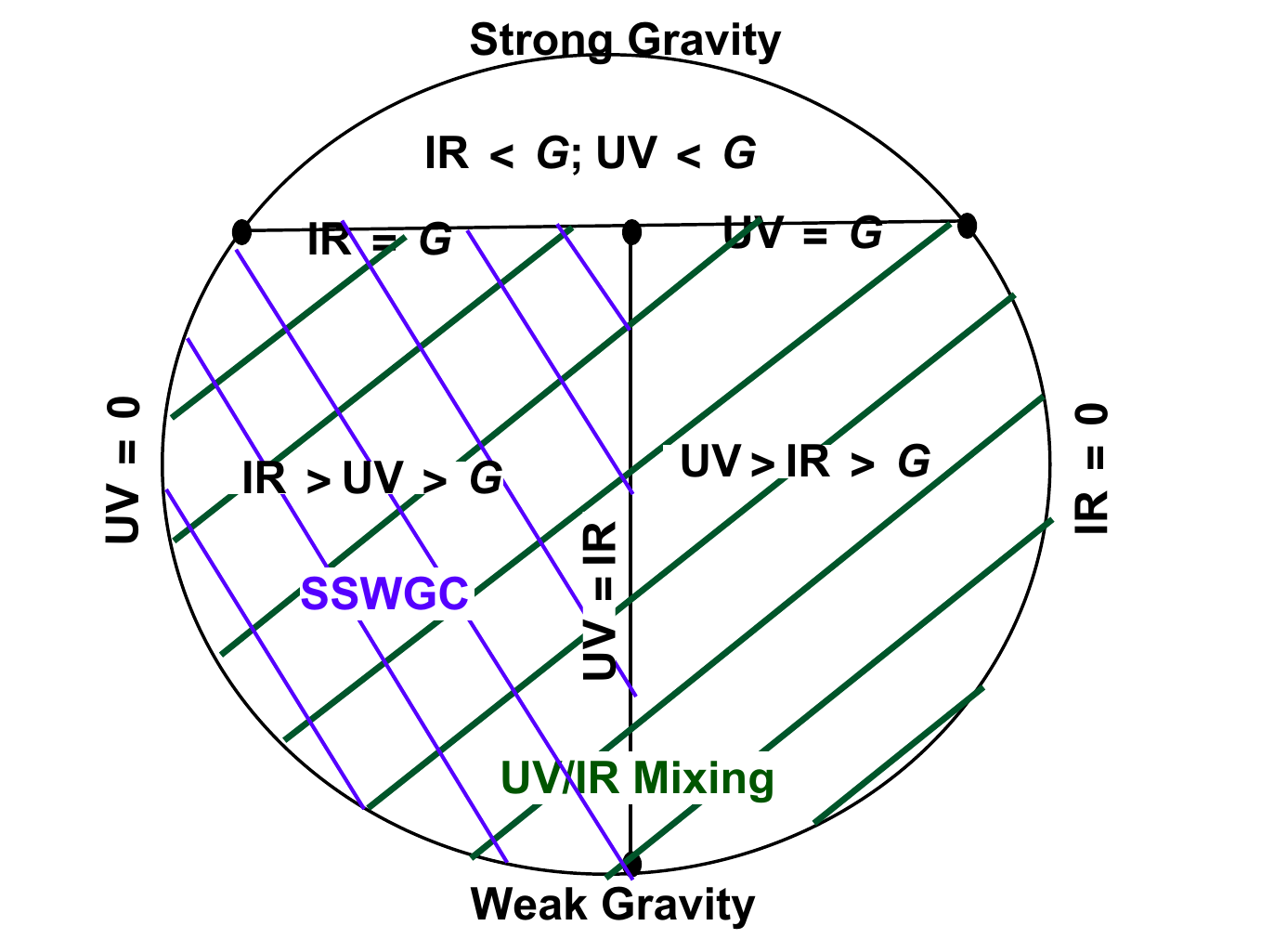} 
  \caption{ The schematic diagram represents the interaction strength of the scalar field presence of weak/strong gravity. The trilinear coupling $\mu$ is related to 
  IR decay and quartic interaction related to UV. Here SSWGC generates tower of light scalar states when $\chi_{S} = 0$ and encapsulates UV/IR mixing
  effects when $\chi_{S} \neq 0$. Where  $\chi_{S} = 2(V''')^{2} -V'' V''''$. }
   \label{fig:planckuvir}
\end{figure}

\begin{figure}[h]
\centering
\includegraphics[width=0.48\textwidth,height=0.38\textwidth]{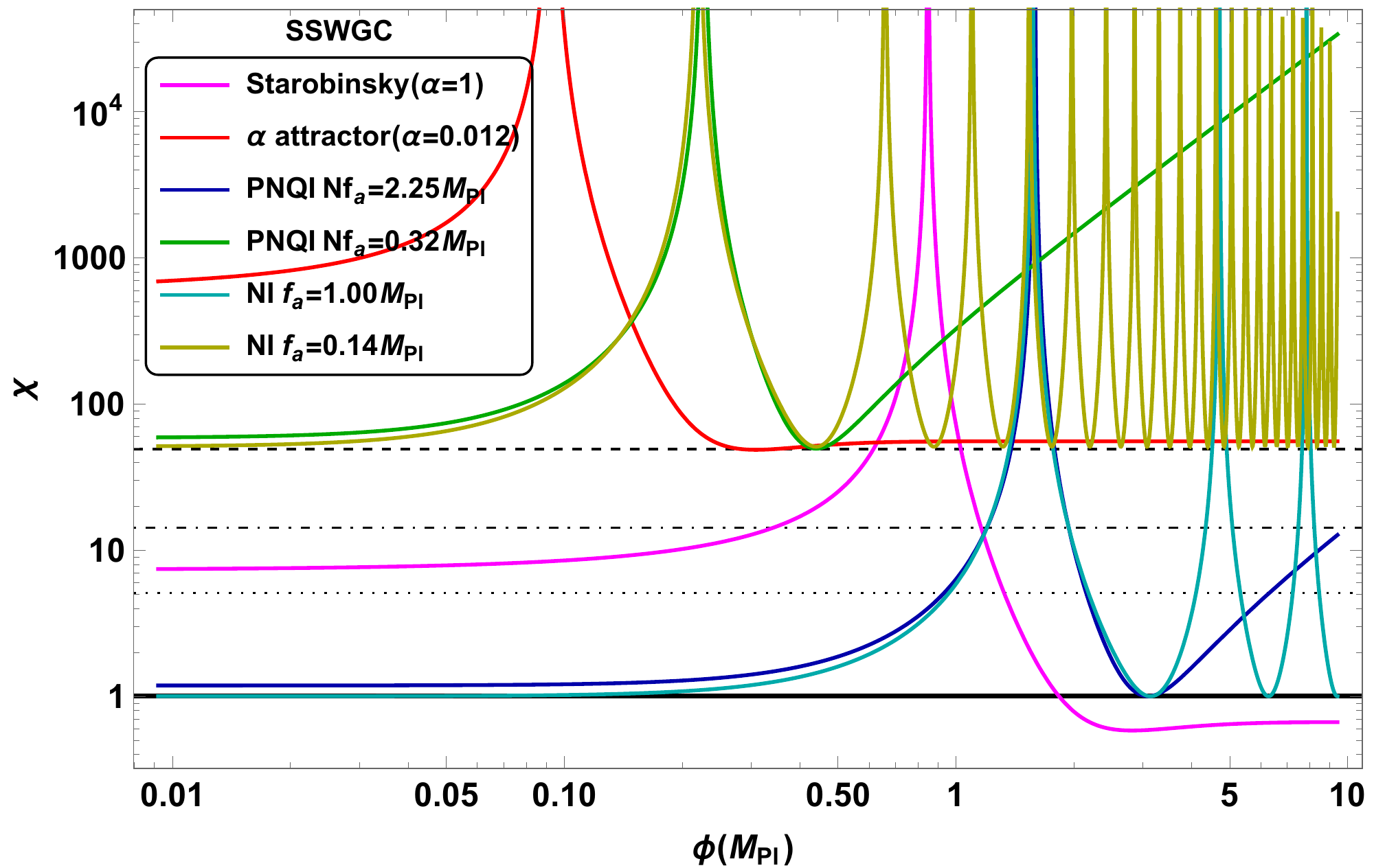} 
  \caption{The strong scalar weak gravity conjecture constraints on Natural inflation, Starobinsky-like($\alpha = 1,\, 0.012$) model and pure natural quintessence inflation.}
   \label{fig:plancksswgc1}
\end{figure}

\noindent  In this section we discuss about an implication of current Planck data on inflationary models particularly Starobinsky, Natural inflation and PNQI models through scalar weak gravity conjecture(SWGC)\cite{wgc1,wgc2}. According to Lyth bound, we know from Planck 2018 result that the scalar field($\phi$) is traverse over the super-Planckian distance. Such field variation urges an infinite tower of states with field dependent mass which is $m(\phi)$ proposed in SWGC. The mass of such WGC scalar($S_{WGC}$) decreases exponentially as a function of scalar field variation as follows \cite{wgc1,wgc2,wgc3}
\begin{equation}
m(\phi+\Delta \phi) \leq m(\phi) e^{-\alpha \frac{\Delta \phi}{M_{\text{Pl}}}} \label{swgc0} 
\end{equation}
Here $\alpha$ is a positive constant $\mathcal{O}(1)$  parameter which is determined by direction of $\Delta \phi$ in the field space. 
The tower of states are nearly massless, such light scalar doesn't need to have interaction with standard model directly, instead they couple through gravitational interaction. Therefore gravity has to be weakest force at the horizon scale.
According to SWGC, the force which is mediated by light scalar must be stronger than graviton mediated force. Such light scalar known as WGC scalar $S_{WGC}$ and the strength can be demonstrated by trilinear coupling ($\mu$) given as
\begin{equation}
 \left(\frac{m}{\mu}\right)^{2} \leq M^{2}_{\text{Pl}} \label{swgc1}
\end{equation}
The trilinear coupling defined as $\mu\equiv\partial_{\phi}m $ and using $m^{2}=V''$ one can translate equation.(\ref{swgc1}) into
\begin{equation}
 \frac{1}{2}(V''')^{2} \geq \frac{(V'')^{2}}{M^{2}_{\text{Pl}}} \label{swgc2}
\end{equation}
But this conjecture applicable only for interaction of WGC scalar and scalar $\phi$, further inequality of the equation.(\ref{swgc1}) emphasize that the existence of the fifth force which is stronger than gravitational force. Such fifth force is excluded by observational constraint on violation of the weak equivalence principle. Further, the SWGC conjecture (equation.(\ref{swgc2})) is inconsistent with the properties of periodic potential such as for axion.
Therefore it has modified in \cite{sswgc}, which is given as follows 
\begin{equation}
 2(V''')^{2} -V'' V''''\geq \frac{(V'')^{2}}{M^{2}_{\text{Pl}}}  \label{sswgc}
\end{equation}
The conjecture equation.(\ref{sswgc}) is valid for any canonically normalized real scalar potential as well as wide range of field value(sub or super-planckian). Here the WGC scalar $S_{WGC}$ is not necessary to satisfy WGC because they play a role in the towers of states when they achieve equality in equation.(\ref{sswgc}). 
Presumably,  absence of light scalar particle(as a mediator), the massive scalar decay through trilinear interaction which is related to $V'''$. This trilinear interaction of scalars can be intuitively thought of the attractive force or appearance of Infra-Red divergence(IR) which is driven by their interaction potential. The additional term $V''''$ related to the quartic coupling of the scalar interaction which strengthened the interaction strength of the scalar against the strength of the gravity compared to equation.(\ref{swgc2}), thus it called a strong scalar weak gravity conjecture(SSWGC). Such inclusion of quartic(repulsive or UV) term $V''\,V''''$ in the SSWGC encapsulates the UV/IR mixing effects \cite{sswgc,uvirphase1,uvirphase2}. The UV/IR mixing possibly would give a better understanding about naturalness problem \cite{craig,selva,seth} in scalar theory and the schematic behavior of UV, IR and UV/IR mixing at weak as well as strong gravity shown in figure.(\ref{fig:planckuvir}) pictorially.
Therefore, SSWGC is applicable for any low energy effective theory with UV completion \cite{sswgc}. One can translate the SSWGC conjecture equation.(\ref{sswgc}) comply to slow roll parameters as follows
\begin{equation}
 \chi \equiv \frac{\xi^{4}_{V}}{\epsilon_{V}\,\eta^{2}_{V}} - \frac{\omega^{3}_{V}}{2\, \epsilon_{V}\,\eta_{V}} \label{exsswgc}
\end{equation}
Where $\chi$ is the order parameter and according equation.(\ref{sswgc}) it must satisfy $\chi \geq 1$ units of $M_{\text{Pl}}$. Here $\epsilon_{V}$ and $\eta_{V}$ are the slow roll parameters of order $n=2$ of Taylor expansion of inflaton potential $V(\phi)$ which are shown in equation.(\ref{srlc1}) and we know that the spectral tilt is defined as $n_{s}=1\,-\,6\epsilon_{V}\,+\,2\eta_{V}$. The order $n=3$ and $n=4$ slow roll parameters are defined in terms of running of $n_{s}$ and running of running of $n_{s}$ parameter $\alpha_{s}$ and $\beta_{s}$ respectively as follows
\begin{figure}[h]
\centering
\includegraphics[width=0.44\textwidth,height=0.36\textwidth]{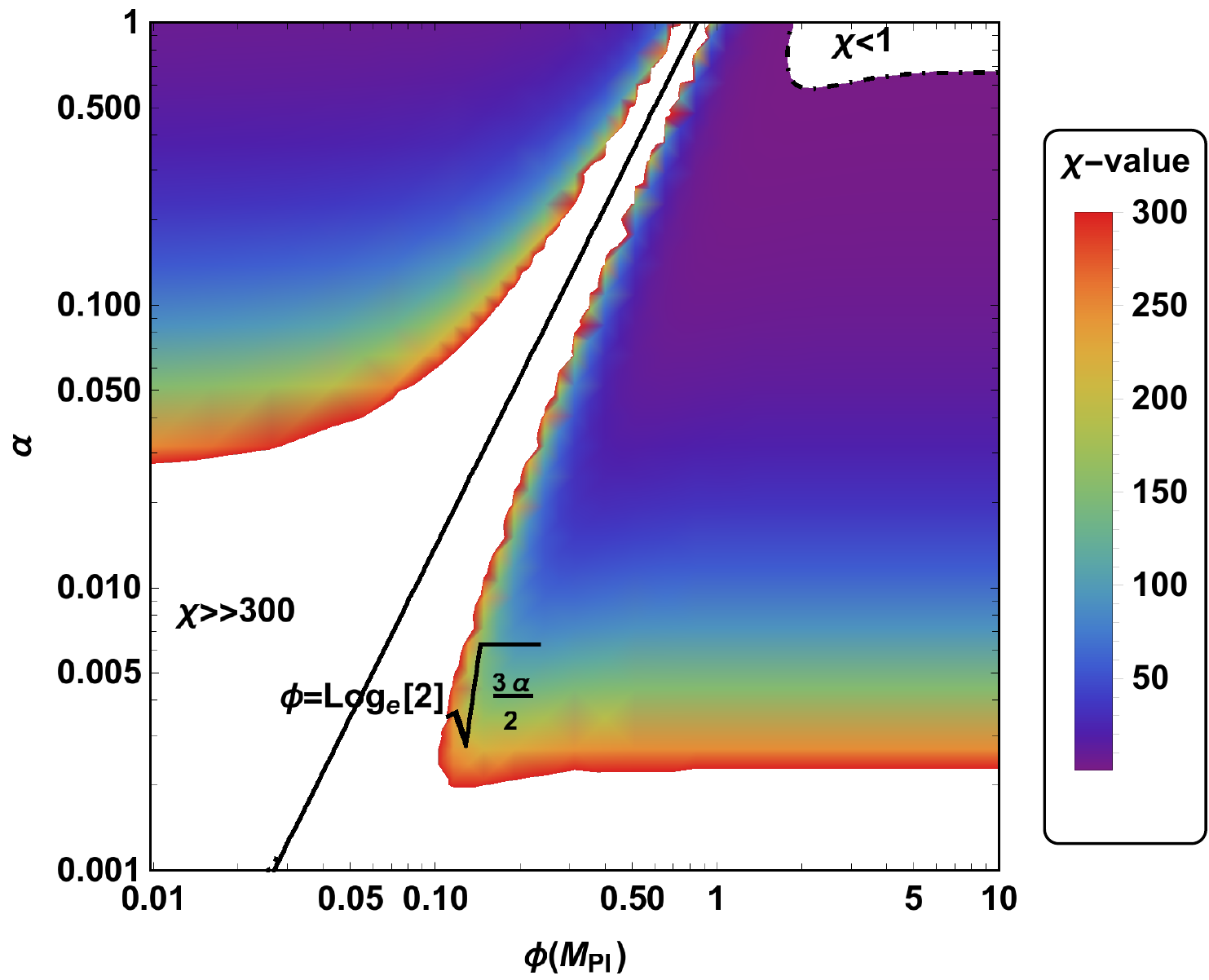} 
 \includegraphics[width=0.4\textwidth,height=0.36\textwidth]{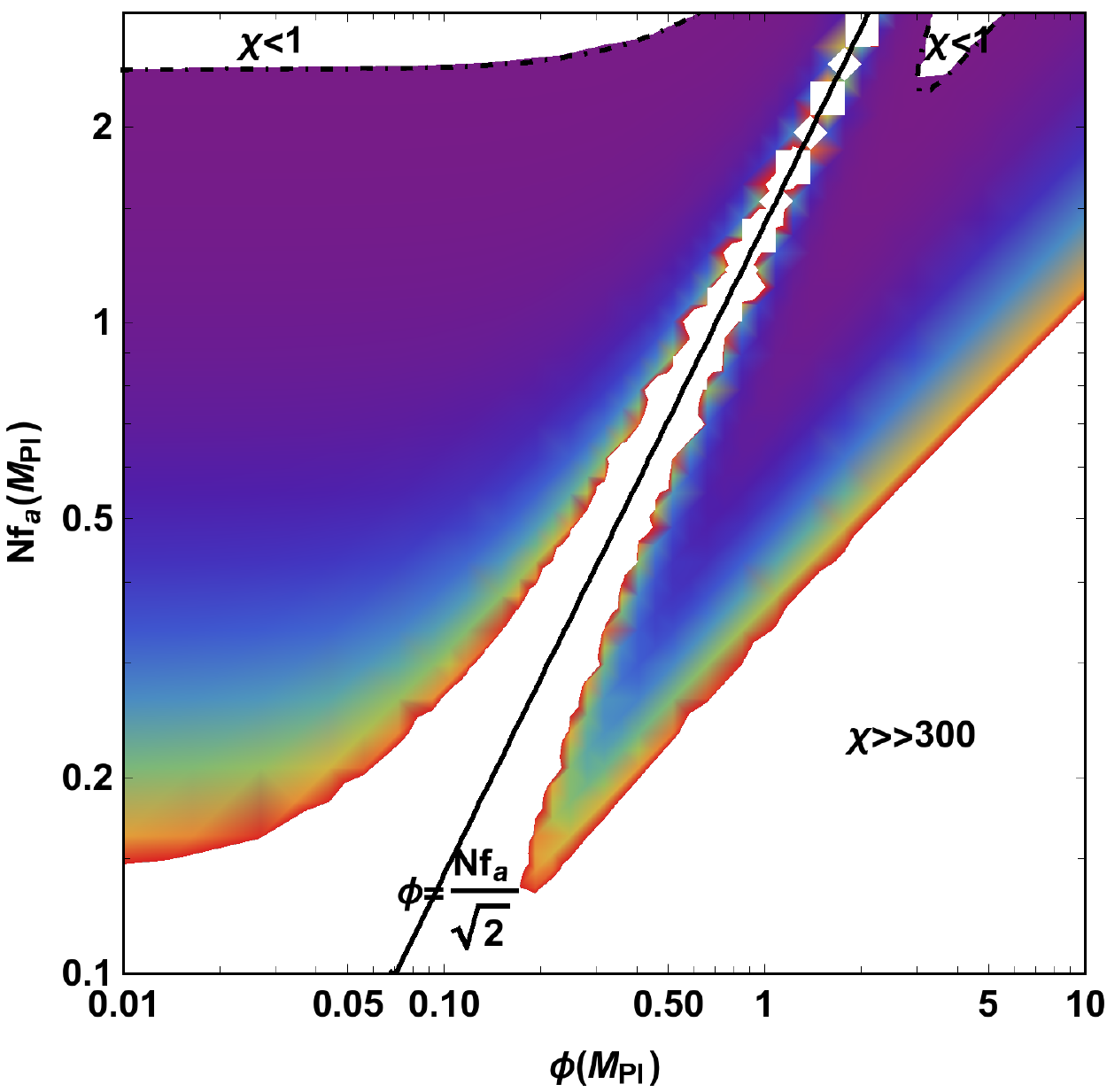} 
  \caption{The figure(left) depicts the strong scalar weak gravity conjecture constraints for Starobinsky-like potential $V=V_{0}\bigg(1-\exp[-\sqrt{\frac{2}{3\alpha}}\frac{\phi}{M_{Pl}}]\bigg)^{2}$ and which diverges 
  at $\phi=\log_{e}[2]\sqrt{3\alpha/2}$ and pure natural quintessence inflation which diverges at axion field $a(x)$ which is $\phi(x)=N f_{a}/\sqrt{2}$(right).}
   \label{fig:plancksswgc}
\end{figure}
 \begin{eqnarray}
 \xi^{2}_{V} & = & \left( \frac{V' V'''}{V^{2}}\right)   =  \frac{1}{2}\left[ \frac{3 r}{16}+n_{s} -1 -\alpha_{s}\right]  \hspace{0.2in} \\
 \omega^{3}_{V} & = & \left( \frac{V'^{2} V''''}{V^{3}}\right) =  \frac{8\beta_{s}-6 r \alpha_{s}+3r^{2}(r-5)}{16} \nonumber \\
  & - &  (\alpha_{s}+\frac{3r}{16} +  2 \xi^{2}_{V})[ 1-\frac{n_{s}}{2}+\frac{r(12r-41)}{64}]
\end{eqnarray}
Here we have taken $\epsilon_{V} = r /16$ and $\eta_{V} =\frac{1}{2}(n_{s}-1+\frac{3r}{8})$. Using \textit{Planck} TT,TE,EE $+$ lowE $+$ BAO $+$ BK14 of Taylor expanded $3$rd and $4$th order slow-roll parameters \cite{planck1},
one can obtain the observational constraints on SSWGC parameter ($\chi$) which must be $\chi \geq 14.27$ and $\chi \geq 49.03$ when it satisfy equation.(\ref{swgc2}) and equation.(\ref{sswgc}) respectively. Here we considered the central value of slow roll parameters given in Planck data.
The constraint $\chi \geq 5.09$ in figure(\ref{fig:plancksswgc1}) represents when slow roll parameter $\epsilon_{4}=0$ with non-zero $\epsilon_{V},\,\eta_{V}$ and $\xi^{2}_{V}$ from \textit{Planck} TT,TE,EE $+$ lowE $+$ lensing $+$ BK14 data 2018 \cite{planck1}.          
The theoretical constraints ($\chi \geq 1$) and observational constraints ($\chi \geq 49.03$) on natural inflation which provides $f_{a} < M_{\text{Pl}}$ and it is true for Axion WGC too. But we have seen that the natural inflation require $f_{a} > M_{\text{Pl}}$ to satisfy $n_{s}-r$ Planck data.

\noindent Similarly Starobinsky inflation is also inconsistent when $\chi \geq 1$ at trans-Planckian field space and it violated SSWGC with observational constraints when $\chi \geq 49.03$. 
Indeed, the Starobinsky-like($\alpha$-attractor) inflation obey the SSWGC when $ \alpha \leq 0.012 $ shown in figure.(\ref{fig:plancksswgc}) and (\ref{fig:plancksswgc1}) which is  allowed within $-2 < \log_{10} \alpha < 4$ given by Planck 2018 result. Thereby, observationally best-fitted Starobinsky-like model($\alpha=1,\,\alpha > 0.012$) inflation would seems facing severe tension by SSWGC not only super-Planckian also sub-Planckian. Therefore such models required large corrections to their potential to intact SSWGC at large as well as small field excursions.

\noindent Our model, PNQI with a large-N limit satisfies the SSWGC with out any correction in it. Here the scalar field replaced by axionic field. Further the property of natural inflation, which is $f_{a} < M_{\text{Pl}}$ has arrived with satisfying the observational data, though the amplitude of $B$-modes would perhaps smaller and smaller as shown in the table.(\ref{tab:qpni}). According to equation.(\ref{exsswgc}) PNQI gets a upper limit on decay constant $f_{a}$ which is $f_{a}\leq 0.32/N$ shown in the figure.(\ref{fig:plancksswgc1}) when $V''>0$. The equation.(\ref{sswgc}) diverges for our model at axion field $a(x)$ which is $\phi(x)=N\,f_{a}/\sqrt{2}$ or in general at $V''=0$. Essentially, SSWGC is applicable for massless as well as massive scalar fields, but background not necessarily to be de Sitter spacetime. Meanwhile if SSWGC is not valid then the string theory motivation towards quantum gravity would get severe tension.

\section{Dynamical dark energy}\label{sec4} 
\noindent After inflation, Universe enters into the (p)reheating era to energize the inflated fields which is required for particle creation. Our interest is to study the early inflation as well as late time inflation (cosmic acceleration) through quintessence axionic field.
Thereby, the usual reheating mechanism doesn't work, instead the instant reheating\cite{instreheatdde1,instreheatdde2,reheatdde1,reheatdde2}, gravitational preheating \cite{gravreheatdde1,chiba1,gravreheatdde2}, Ricci reheating \cite{riccireheatdde1,riccireheatdde2} etc., would be an alternative mechanism for reheating. Though the quintessence reheating mechanism makes important consequences on the model parameters at reheating era but one can achieve proper initial conditions for late time acceleration which followed by same inflationary model parameters, therefore we will discuss about reheating mechanism in future which is beyond the scope of this paper.

\noindent However, the quintessence scalar field model is a natural candidate to unify the inflation and dark energy but it faces the challenge due to quantum corrections which spoil the flatness of the potential and fifth force problem by ultra-light scalar. Axionic models solve such problems by shift symmetry even though when symmetry broken in a controlled manner to get a axion naturally light. Axionic derivative couplings suppresses the fifth-force constraints by nature \cite{carrol}. Such kind of axionic potential drives early time as well as late time inflation by pseudo-Nambu Glodstone Boson(pNGB) which is slowly rolling on curvature of the potential. Here the energy density of the quintessence field is not necessarily constant, indeed it varies dynamically as the universe evolves with respect to time \cite{dde2,dde4}. Further, quintessence field has to survive after inflation to till date to satisfy the property of dynamical dark energy \cite{dde1,dde2,dde3}. 
%

\noindent The quintessence is described by a minimally coupled scalar field $\phi$ with the potential $V(\phi)$. The action for quintessence field with nonrelativistic matter in the presence of barotropic perfect fluid can be written as
\begin{equation}
 S = \int dx^{4} \sqrt{-g} \left[ \frac{1}{2}R-\frac{1}{2}g^{\mu \nu} \partial_{\mu}\phi \partial_{\nu}\phi-V(\phi)\right] + S_{m}
\end{equation}
Here $S_{m}$ is the action for nonrelativistic matter and $g$ is the determinant of FLRW metric $g_{\mu\,\nu}$ with the scale factor $a(t)$. Therefore to avoid confusion among the axion field($a(x)$) and scale factor($a(t)$), we follow the notation $\phi(x)$ for axion field and for scale factor as it is. The dynamical equations of motion are written as
\begin{eqnarray}
\ddot{\phi} & + & 3H\dot{\phi}+V'(\phi) =  0 \label{euler} \\ 
3H^{2} & = & \rho_{\phi} + \rho_{m} \\
\dot{\rho}_{m} & + & 3 H\rho_{m} = 0
\end{eqnarray}
Here $H=\dot{a}/a$ and $V'(\phi)=dV/d\phi$. Where $a$ is scale factor and dot denotes derivatives with respect to time. In addition to that 
\begin{equation}
 H^{2}=\left(\frac{\dot{a}}{a}\right)^{2}=\frac{\rho}{3}\quad \text{and} \quad \frac{\ddot{a}}{a} =-\frac{(\rho+3P)}{6} \label{hub}
\end{equation}
The equation of state $\omega$ for scalar field reads
\begin{equation}
 \omega=\frac{P_{\phi}}{\rho_{\phi}} = \frac{\dot{\phi}^{2}/2 - V(\phi)}{\dot{\phi}^{2}/2 + V(\phi)}
\end{equation}
Here $P_{\phi}$ and $\rho_{\phi}$ are the spatially-averaged (background) pressure and density of the scalar field respectively. The pressure and density are follows relation  at radiation, matter and dark energy dominated era are $P_{\phi} = -\rho_{\phi}/3 ,\, P_{\phi} =0$ and $P_{\phi} \approx -\rho_{\phi}$  respectively. In order to deal with cosmological dynamics, we can introduce following dimensionless variables such as
\begin{equation}
 x  =  \frac{\phi_{,N}}{\sqrt{6}}; \quad y =  \sqrt{\frac{V(\phi)}{3H^{2}}}; \quad \lambda  = -\frac{V'}{V} \label{deeq} \\
\end{equation}
Where $N=\ln a$ and $\phi_{,N}=d\phi/d N$. Assume that scalar field (Dark energy) and matter are only presented then $\Omega_{\phi}+\Omega_{\text{m}}=1$.
The scalar field (Dark energy) density parameter $\Omega_{\phi}= \rho_{\phi}/3H^{2}$ can be written as
\begin{equation}
 \Omega_{\phi}= x^{2}+y^{2}
\end{equation}
and for calculational simplicity, one can introduce the parameter $\gamma$ which is 
\begin{equation}
 \gamma =1+\omega =\frac{2x^{2}}{x^{2}+y^{2}}
\end{equation}
here we work on where $\omega$ is near to $-1$, Which means $\gamma = \dot{\phi}^{2}/\rho_{\phi} <<1 $. Therefore, in order to
find the fixed points of $x$ and $y$ for the system which satisfies equation.(\ref{euler}) and (\ref{hub}) we set $dx/dN=0$ and $dy/dN=0$. But $x$ and $y$ are related
to the observable quantities $\Omega_{\phi}$ and $\gamma$, therefore we assume that $dx/dN>0$ then we get the evolution equation for $\Omega_{\phi}$, $\gamma$ and $\lambda$, which are \cite{chiba,saridakis,scherrer1,rrangarajan}
\begin{eqnarray}
 \frac{d\gamma}{dN} & = & -3\gamma(2-\gamma)+\lambda(2-\gamma)\sqrt{3\gamma \, \Omega_{\phi}} \\
 \frac{d\Omega_{\phi}}{dN} & = & 3(1-\gamma)\Omega_{\phi}(1-\Omega_{\phi})\\
 \frac{d\lambda}{dN} & = & -\sqrt{3} \lambda^{2}(\Gamma-1) \sqrt{\gamma \, \Omega_{\phi}} \\
 \text{Where} & & \Gamma = \frac{V V''}{V'^{2}} 
\end{eqnarray}
As mentioned earlier, the limit $\gamma << 1$ produces the solution for $\Omega_{\phi}$ when we take $\rho_{\phi}\approx \rho_{\phi_{0}} \approx V(\phi_{0})$ at present, which is given by
\begin{equation}
 \Omega_{\phi} = [1+(\Omega_{\phi_{0}}^{-1}-1)a^{-3}]^{-1} \label{engden}
\end{equation}
Where $\Omega_{\phi_{0}}$ is the present value of $\Omega_{\phi}$. Importantly, the dynamical dark energy models are classified into thawing and freezing \cite{caldwell} 
and which are characteristically has different behaviour. Depends on value of the $\Gamma$, we can classifies them as when $\Gamma > 1$ known as Tracker freezing model \cite{dde3}, 
when $\Gamma = 1$ Scaling freezing model \cite{chiba} and when $\Gamma < 1$ called as Thawing frozen model \cite{scherrer1,rrangarajan}. In ref.\cite{chiba,sugiy}, the observational constraints on quintessence dark energy model has explained in detail. In the case of tracking and scaling model, the equation of state gradually slowing down when the field rolling down along the potential and they enters into the acceleration phase when they satisfy tracking and scaling condition respectively. Importantly, the simple exponential potential doesn't start acceleration with observational constraints, to alleviate this problem one requires double exponential potential in the scaling model\cite{chiba,sugiy}. In general, the tracking and scaling field freeze after the inflation and they doesn't have enough energy to start the late time acceleration. Instead of modifying quintessence potential, one can introduce non-minimal coupling of the neutrino with the background scalar field, then the mass varying neutrino (MaVaN) provide the energy to scalar field to start acceleration with simple exponential scaling potential as well as power-law tracking potential \cite{Hossain1,Hossain2,sami}. Therefore the non-relativistic neutrino makes significant consequences on equation of state and energy density of dark energy of scaling and tracking models.

\noindent The thawing model of the quintessence field $\phi$ is nearly frozen in early time due to Hubble friction. In this epoch, the mass of the quintessence field $m_{\phi}$ is lesser than Hubble friction then the equation of the state $\omega$ is close to $-1$ which ensures the approximation $\gamma \ll 1$ in the thawing model i.e $\dot{\phi}^{2}\ll V(\phi)$ \cite{scherrer1}. Later, the equation of state departs from $-1$ when the Hubble friction drops below or equal to $m_{\phi}$. This is achieved when $\lambda \neq 0$, which provides the unstable solution of the equation.(\ref{deeq}) \cite{chiba}. The thawing model almost behave like cosmological constant in most of the acceleration epoch. Unlike scaling and tracking model, the thawing do not have significant restriction on $\omega_{0}$ and $\Omega_{\phi_{0}}$ parameters when we introduce non-minimal neutrino coupling to the quintessence scalar field. Because thawing model depends on three parameters $\omega,\,\Omega_{\phi}$ and $K$, which makes the thawing DE analysis as more general. Here the parameter $K$ doesn't restricted by mass of the non-relativistic neutrino\cite{sugiy}. Our model follows thawing quintessence, therefore we will explain more detail about thawing in the next section.

\subsection{Thawing quintessential axion}
\begin{figure}[h]
\centering
\includegraphics[width=0.46\textwidth,height=0.34\textwidth]{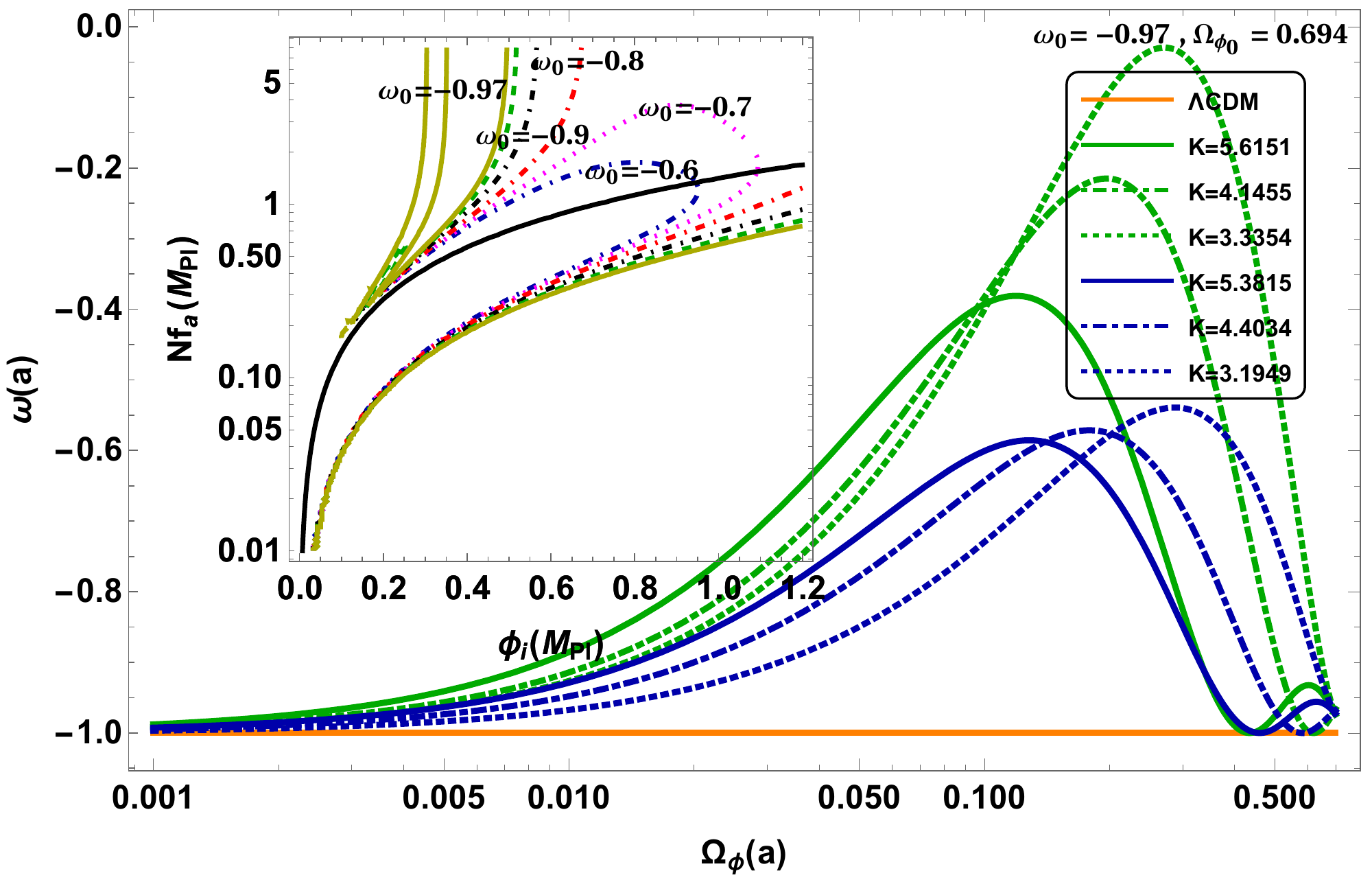} 
\includegraphics[width=0.46\textwidth,height=0.34\textwidth]{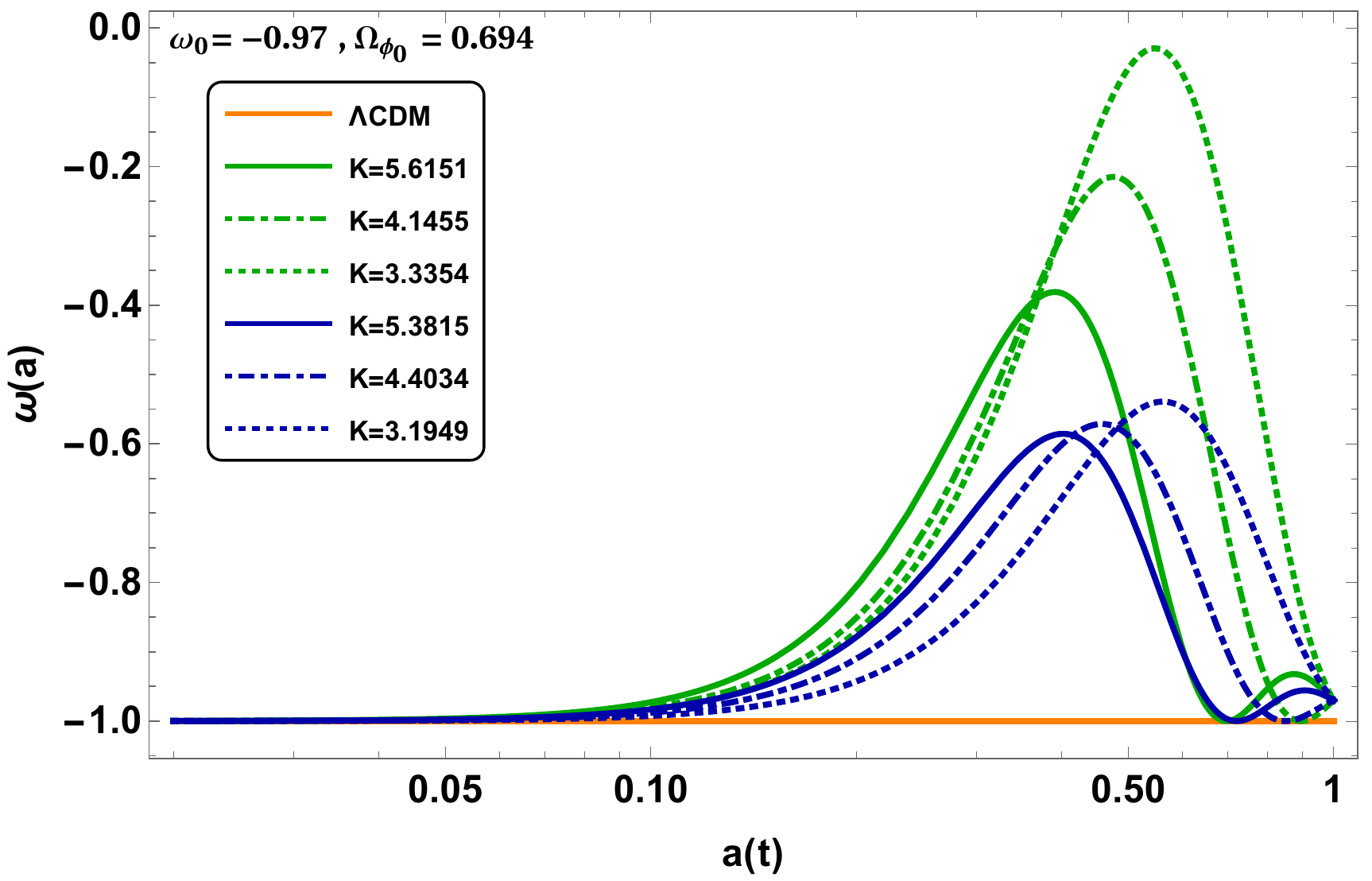} 
  \caption{Inset of left figure depicts the initial field value for fixed $\omega_{0}$ and the favored region constrained by $\phi_{i} < Nf_{a}/\sqrt{2}$(above black line). Equation of state $\omega(a)$ evolves as energy density parameter($\Omega_{\phi}$) of the quintessence field evolves 
  from equipartition at $a\approx0$(end of inflation) to $a=1$(present). Green and blue curves shows such evolution happens when $Nf_{a}=0.4\,M_{Pl}$ and $Nf_{a}=1.0\,M_{Pl}$ for fixed $\omega_{0}$, $\Omega_{\phi_{0}}$ and different initial field value(oscillatory) respectively.
  }
   \label{fig:dde}
\end{figure}

\noindent  In this quintessence model, the field is nearly frozen far away from the minimum of the potential by Hubble friction at early time epoch of the cosmic acceleration. Later, the field starts to roll slowly towards potential minimum due to diminishing of the Hubble friction and such damping declined to $H_{0}$ today and we get energy density parameter $\Omega_{\phi_{0}}$ at $\phi_{0}$(present field value). 
\noindent Here we study pseudo-Nambu-Goldstone boson(PNGB) as thawing quintessence and the potential as proposed earlier in section.\ref{sec2}: 
\begin{equation}
 V(\phi) = M^{4}  \left\{1- \exp\left[-\left(\frac{\phi}{N f_{a}}\right)^{2}\right] \right\}
\end{equation}
Here $\phi$ and $f_{a}$ are the axion and axion decay constant respectively. The quintessence axion rolls slowly on potential from initial field value $\phi=\phi_{i}$ towards local minimum of the potential until the Hubble $H$ drops to mass of the axion $m_{a}(\phi)$ (field dependent). Note that, here the initial field $\phi_{i}$ is nothing but field value of axion when the inflation ends. After that the field oscillate around the local minimum with amplitude diluting as $\sim a^{-3/2}$.
Therefore one can substitute change of variable $\phi(t) = a^{-3/2} u(t)$ in equation.(\ref{euler}) and evaluate the potential around initial value $\phi_{i}$ which is up to second order in Taylor expansion then we get 
\begin{equation}
 \ddot{u}-k^{2}u\simeq -a^{3/2}V'(\phi_{i})
\end{equation}
Here $k = \sqrt{3V(\phi_{i})/4-V''(\phi_{i})}$ and with the limit $\gamma << 1$ the scale factor approximated from $\Lambda$CDM model which is 
\begin{equation}
 a(t)=\left(\frac{1-\Omega_{\phi_{0}}}{ \Omega_{\phi_{0}}}\right)^{1/3} \sinh^{2/3}(t/t_{\Lambda})\quad \text{at} \quad t_{\Lambda} = 2/\sqrt{3V(\phi_{i})} \label{scale}
\end{equation}
Then the solution for $\phi$ with initial conditions $\phi(0)=\phi_{i} \, , \, \dot{\phi}(0)=0$ and $V''(\phi_{i})\neq 0 $, we get
\begin{equation}
 \phi(t) = \phi_{i}+\frac{V'(\phi_{i})}{V''(\phi_{i})}\left( \frac{\sinh(k t)}{k t_{\Lambda}\sinh(t/t_{\Lambda})} -1\right) \label{phisol}
\end{equation}
Sine we assume the initial energy density of field is $\rho_{\phi_{i}} = V(\phi_{i})$ then the approximated equation of state $1+\omega\simeq \dot{\phi}^{2}/V(\phi_{i})$ which can be written as follows(known as \textit{Scherrer-Sen} equation of state \cite{scherrer1,scherrer2,scherrer3})
\begin{equation}
 \omega(a) = -1 + (1+\omega_{0})a^{-3} \mathcal{F}(\Omega_{\phi}) \label{schersen}
\end{equation}
From equation.(\ref{engden}) and (\ref{scale}) we get
\begin{eqnarray}
a(t) & = & \left(\frac{\Omega_{\phi_{0}}(1-\Omega_{\phi})}{\Omega_{\phi}(1-\Omega_{\phi_{0}})}\right)^{-1/3} \\
\mathcal{F}(\Omega_{\phi})  & = & \left[ \frac{K \cos(\frac{K t}{t_{\Lambda}}) - \frac{1}{\sqrt{\Omega_{\phi}}}\sin(\frac{K t}{t_{\Lambda}})}{K \cos(\frac{K t_{0}}{t_{\Lambda}}) - \frac{1}{\sqrt{\Omega_{\phi_{0}}}}\sin(\frac{K t_{0}}{t_{\Lambda}} )}\right]^{2} \\
\frac{t_{0}}{t_{\Lambda}} & = & \tanh^{-1}(\sqrt{\Omega_{\phi_{0}}}) \\
\frac{t}{t_{\Lambda}} & = & \sinh^{-1}\left(\sqrt{\frac{\Omega_{\phi}}{1-\Omega_{\phi}} }\right) 
\end{eqnarray}
Importantly the potential dependent parameter defined when $K^{2}>0$ and $V''(\phi_{i}) > 0$ is
\begin{equation}
  K  =  \sqrt{\frac{4V''(\phi_{i})}{3 V(\phi_{i})}-1} \label{kvalue}
\end{equation}
The planck result \cite{planck2} suggest that the dark enegy is almost act like cosmological constant or thawing quintessence which implies that $\omega_{0} $ is not exactly equal to $-1$ therefore $\gamma \ll 1$. However, the parameter $\gamma$ is inversely proportional to $V(\phi)$ but $\dot{\phi}^{2}$ is very small than $V(\phi_{0})$ at present which validate the approximation. Here the $\dot{\phi}$ at present $\phi_{0}$ can be calculated from equation.(\ref{phisol}). Eventually, $\gamma$ follows Scherrer-Sen relation (\ref{schersen}) always, which satisfy the observational data.
Therefore we fixed the value $\omega_{0} = -0.97$ within the error limit and the present value of dark energy density $\Omega_{\phi_{0}} = 0.694$ for our analysis. We have arrived three initial field values which is due to oscillatory field solution and it respects $V''>0$ as well as $\phi_{i} < Nf_{a}/\sqrt{2}$, which is required for cosmological evolution towards local minimum of the potential as shown in figure.(\ref{fig:dde}).\\
\noindent The initial field value of thawing analysis and approximations are carried out by taking the same inflationary parameters and same potential thereby  known as quintessence inflation model. In order to follow lower value of tensor-scalar ratio ($r$), we have chosen $Nf_{a}$ from table.(\ref{tab:qpni}) which are $0.4\,M_{Pl}$ and $1.0\,M_{Pl}$.
The Scherrer-Sen equation of state(equation.(\ref{schersen})) for our model follows sine and cosine functions instead of hyperbolic functions by definition of $K$ given in equation.(\ref{kvalue}).
Here the value of $K$ obtained from initial field value at end of the inflation which sets the initial conditions for dynamical scalar energy density($\rho_{\phi}$) as well as equation of state of the dark energy component($\omega_{\phi}$).

\noindent The initial field value($\phi_{i}$) computed when the solution of the equation.(\ref{schersen}) has solved based on present $\omega_{0}$ and $\Omega_{\phi_{0}}$ as a input. 
At high redshift nearly $z=10^{28}$ (where $a^{-1}=1+z$) with the initial field value $\phi_{i}$, the energy density parameter $\Omega_{\phi_{i}}$ taken to be $10^{-4}$ when all the components were in equipartition state. Since the thawing models naturally attain the value of the equation of state close to $-1$ at initial field value, then they start to increase when the field rolls towards the potential minimum which is near lower redshift as shown at the figure.(\ref{fig:dde}). Our result doesn't affect the BBN by thawing quintessence because of they have negligible energy density parameter $\Omega_{\phi}$ at $t_{\text{BBN}}$ and $\omega_{\phi}=-1$ at high red shift.
Current value of dark energy density $\rho_{\phi_{0}} = \rho_{c}\, \Omega_{\phi_{0}} $ is equal to $(2.25\times 10^{-3} eV)^{4}$ where critical density $\rho_{c}=3.67\times 10^{-47}\,\text{GeV}^{4}$ and $ \Omega_{\phi_{0}}=0.694$, 
in order to achieve these values, potential($V(\phi)$) has to meet same value for particular field value($\phi_{0}$) which need not to equal to zero.
In principle, we can get explicit symmetry breaking scale($\Lambda$) nearly reduced Planck scale ($M_{\text{Pl}}=2.4\times10^{18}GeV$), at the same time we follow $M<F_{a}$ thus the height of the potential($V(\phi_{i})=\rho_{\phi_{i}}$)
at initial field $\phi_{i}=0.195\,M_{\text{Pl}}$, $M=10^{-2}\,M_{\text{Pl}}$ and $Nf_{a}=F_{a}=0.4\,M_{\text{Pl}}$ we get  $\rho_{\phi_{i}}=7.02\times10^{64}\,GeV^{4}$.
These values are taken from figure.(\ref{fig:dde}) particularly $K=5.6151$. Eventually it is warping of height of the potential by cosmic acceleration, further sake of fine tuning \cite{brax,dodelson} we infer the $\phi_{0}$ value is 
approximately $10^{-56}\,F_{a}$ with breaking scale $\Lambda$ remains same and it can be $\Lambda \leq 2\times10^{-3}\,M_{\text{Pl}}$ when $N\geq4$. The mass of the quintessence particle $m_{a}$ at $\phi_{0}$ is equal to
$\sqrt{2\,\rho_{\phi_{0}}}/F_{a}\,\simeq \,7.5\times10^{-33}\,eV$. This quintessence axion could be dark energy candidate with mass which is ultra-light.

\section{Conclusion}
\label{conclusion}
\noindent In this work we have proposed the pure natural quintessence inflation model in accord with Witten's large-N limit conjecture. Our model satisfies current Planck 2018 results and valid for  very low tensor-scalar ratio as well, which leads sub-planckian cosmic universe. Further we have discussed SSWGC with higher order slow-roll parameters which sets the criteria whether the model is valid in both sub-panckian as well as super-planckian field space which is applicable for any spacetime(de-sitter or quasi de-sitter). We have noticed that Starobinsky model require large quantum corrections to satisfy sub-planckian and also for super-planckian. Similarly natural inflation lost their natural property which means that the spontaneous breaking scale has to be lesser than planck scale when they satisfy Planck result. In the late time evolution, in our model, the quintessence field evolves with respect to cosmic time from initial field value($\phi_{i}$) to present non-zero minimum field value($\phi_{0}$) i.e $10^{-56}\,F_{a}$ which required to achieve present vacuum energy density $\rho_{\phi_{0}} = (2.25\times 10^{-3} eV)^{4}$. Such required minimum field obtained when the spontaneous breaking scale $F_{a}=Nf_{a}$ is equal to $0.4\,M_{\text{Pl}}$ and explicit breaking scale $M=10^{-2}\,M_{\text{Pl}}$ which maintain the property of natural scale inflation as well as present vacuum energy density. 

\acknowledgments
\noindent Author would like to acknowledge Prof. Partha Konar (PRL Ahmedabad) and Prof. Debasish Majumdar (Saha Institute of Nuclear Physiscs Kolkata) for their valuable discussions and comments. This work is supported by Physical Research Laboratory (PRL), Department of Space, Government of India and also acknowledge the computational  support from Vikram-100 HPC at PRL.

\appendix
\section{}
\subsection{ Axion potential}
\label{app:poten}
\noindent In the pure Yang-Mills theory, the instanton may not generate cosine type potential because such weak coupling instanton computation breaks-down for strongly coupled scenario. In the large-$N$ limit, the gluon loops are contribute to the vacuum energy, therefore vacuum energy is of order of $N^{2}$. According to Witten's conjecture
$$ E(N,\theta) = N^{2} h(\theta/N) \quad = \, N^{2}\Lambda^{4} g(\theta/N)$$
From equation(\ref{lagra1}) we can write
\begin{equation}
 \frac{d^{2}E(\theta)}{d\theta^{2}}|_{\theta=0} = \left(\frac{1}{32\pi^{2}}\right)^{2}\int d^{4}x \, \langle T[(N F_{\mu\nu}\widetilde{F}^{\mu\nu})(x_{1}).(N F_{\mu\nu}\widetilde{F}^{\mu\nu})(x_{0})]\rangle
\end{equation}
Here $F\widetilde{F}$ also has same color structure as $FF$ \cite{NNI}. Suppose, correlator has infrared poles then 
$$ F_{\mu \nu} \widetilde{F}^{\mu \nu} = i q_{\mu} K^{\mu}\rightarrow0 \quad (\text{When }\,q_{\mu}\rightarrow 0).\quad \text{But the correlators of}\, K^{\mu}\, \text{is}\quad \langle K^{\mu}K^{\nu} \rangle \sim \frac{\eta^{\mu \nu}}{q^{2}}$$
$$ \langle q_{\mu}K^{\mu} q_{\nu}K^{\nu} \rangle \sim 1 \quad \text{then we get } \quad \frac{d^{2}E(\theta)}{d\theta^{2}}|_{\theta=0} \approx N^{2} $$ 
In the case of axion, substitute $\theta=a(x)/f_{a}$ then
$$ \frac{d^{2}V(a)}{d a^{2}}|_{a=0} = m^{2}_{a} \quad \text{which demands that} \quad \frac{d^{2}g}{d a^{2}}|_{a=0} \approx \frac{1}{N^{2}f^{2}_{a}}$$
Here the double derivative of the potential evaluated at $a(x)=0$, which
defines that quadratic term in the potential proportional to the mass term. 
\begin{figure}[h]
\centering
\includegraphics[width=0.5\textwidth,height=0.3\textwidth]{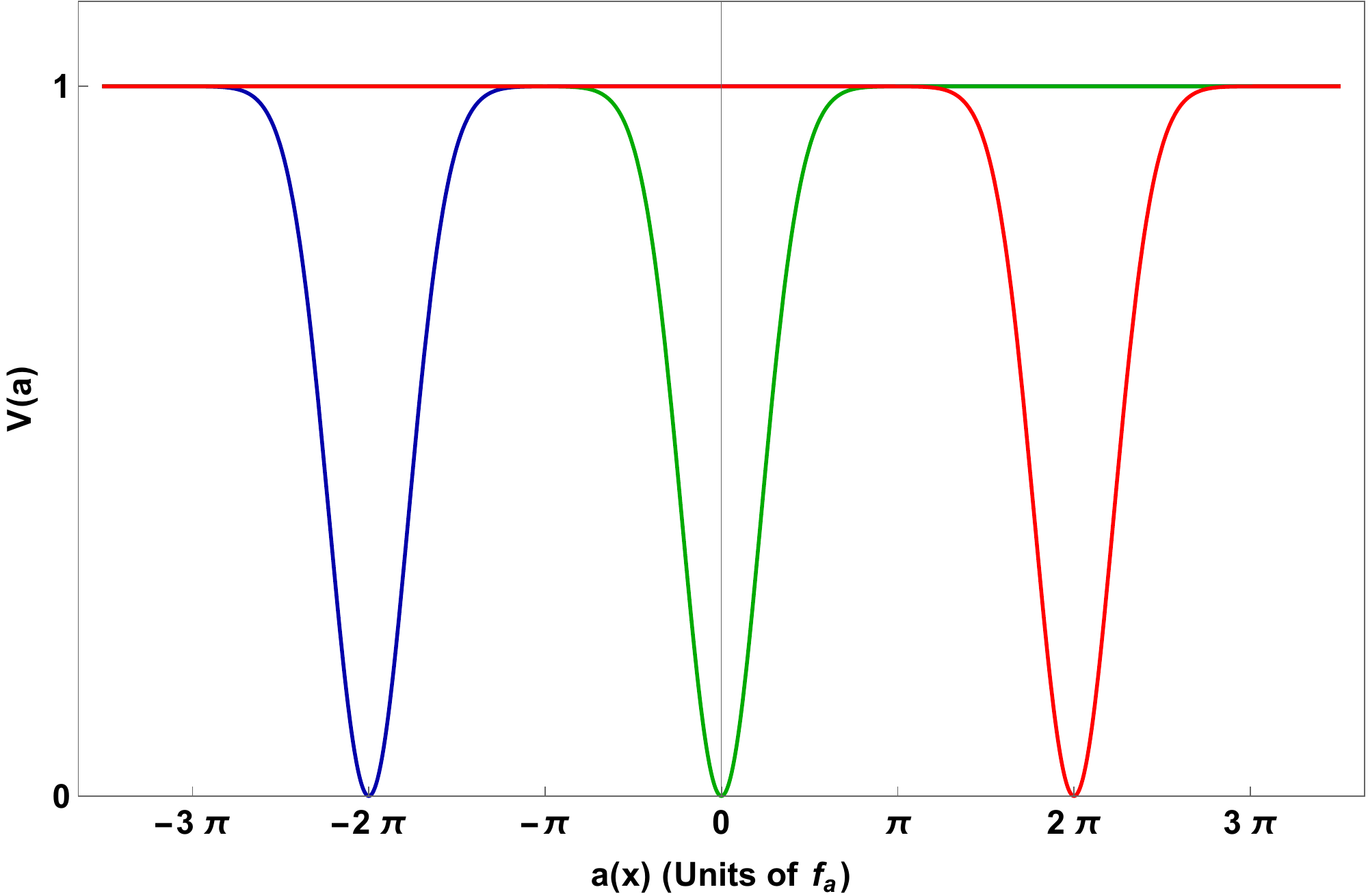} 
 \caption{ Potential with the periodicity of $2\pi f_{a}$ }
 \label{fig:pot}
   \end{figure}
   \noindent Therefore we get $m^{2}_{a} \approx \Lambda^{4}/f^{2}_{a}$ and the potential should be $\mathcal{O}((a/Nf_{a})^{2})$ or Taylor expansion of the square of the field which satisfies CP property with periodicity $a(x)= a(x)+2\pi Nf_{a}$. Thus we can conjecture the potential as follows
\begin{equation}
 V(a) = M^{4}\left( 1- \sum_{n=0}^{\infty}(-1)^{n}\, C_{n}\, \left(\frac{a(x)}{Nf_{a}}\right)^{2n}\right)
\end{equation}
    Where $M= \sqrt{N}\, \Lambda$ and one can assume $C_{n}=\frac{1}{n!}$. The Taylor expansion of the potential converges similar to Cosine potential and it follows quadratic for large-$N$ limit. Importantly, at the deconfinement phase transition, the potential will become weaker with increasing value of field which results that the potential is extremely flat at large filed value. Therefore the potential takes the form as follows and it has plotted as figure (\ref{fig:pot})
   \begin{equation}
    V(a) = M^{4}( 1- \exp\left[-(a(x)/Nf_{a})^{2}\right]
   \end{equation}
   \subsection{ Slow-roll condition}
   \noindent First slow-roll parameter:
    \begin{equation}
   \epsilon = \frac{1}{2}\left(\frac{V'}{V}\right)^{2} = \frac{1}{2} \left\{ \frac{2 a(x)}{(Nf_{a})^{2}} \left[ \exp[(a(x)/Nf_{a})^{2}]-1\right]^{-1} \right\}^{2}  
    \end{equation}

\noindent Second slow-roll parameter:
 \begin{equation}
  \eta = \frac{V''}{V} = \frac{2}{(Nf_{a})^{2}} \left[ \exp[(a(x)/Nf_{a})^{2}]-1\right]^{-1}\left[ 1-2\left(\frac{a(x)}{Nf_{a}}\right)^{2}\right]
 \end{equation}
\noindent Required amount of e-folds:
 \begin{equation}
  N_{e} = \frac{(Nf_{a})^{2}}{2} \left\{ \frac{1}{2} \left( Ei\left[\left(\frac{a(x)_{\text{ini}}}{Nf_{a}}\right)^{2}\right]-Ei\left[\left(\frac{a(x)_{\text{end}}}{Nf_{a}}\right)^{2}\right]\right)+ \ln\left( \frac{a(x)_{\text{end}}}{a(x)_{\text{ini}}}\right)\right\}
 \end{equation}
Where $a(x)_{\text{ini}/\text{end}}$ are the initial and end field value of the inflaton respectively. $Ei(\phi)$ is the exponential integral
\begin{equation}
 Ei(\phi) = -\int^{\infty}_{\phi} \frac{e^{-t}}{t}\, d\,t
\end{equation}
Series form of $Ei(\phi)$ (Ramanujan's faster convergent series):
\begin{equation}
 Ei(\phi) = \gamma_{E} + \ln(\phi)+\exp(\phi/2)\left(\sum_{n=1}^{\infty} \frac{(-1)^{n-1}\phi^{n}}{n!2^{n-1}} \sum_{k=0}^{[(n-1)/2]}\frac{1}{2k+1}\right); \quad \phi>0 
\end{equation}
\subsection{ Slow-roll SSWGC condition}
\noindent Strong Scalar Weak Gravity Conjecture: \\
From equation(\ref{sswgc})
\begin{equation}
2\left(\frac{V'''}{V''}\right)^{2} - \left(\frac{V''''}{V''}\right)  \geq  \frac{1}{M^{2}_{\text{Pl}}}  
\end{equation}
Using the slow roll parameter up to fourth-order
\begin{equation}
 \epsilon_{V}= \frac{M_{\text{Pl}}^{2}}{2} \left(\frac{V'}{V}\right)^{2}, \quad \eta                                                                                                                                                                                                                                                                                                                                                                                                                                                                                                                                                                           _{V}= M_{\text{Pl}}^{2} \left(\frac{V''}{V}\right), \quad \xi^{2}_{V} = M_{\text{Pl}}^{4} \frac{V''' V'}{V^{2}} \quad \text{and} \quad \omega^{3}_{V} = M_{\text{Pl}}^{6} \frac{V'''' V'^{2}}{V^{3}}
\end{equation}
We get
\begin{equation}
 \chi(\phi) \equiv \frac{\xi^{4}_{V}}{\epsilon_{V}\,\eta^{2}_{V}} - \frac{\omega^{3}_{V}}{2\, \epsilon_{V}\,\eta_{V}} \geq 1
\end{equation}
The cosmological observations satisfies that $\chi > 1$ meanwhile $\chi_{obs}$ get a lower bound by upper bound of the tensor-scalar ratio ($r<10^{-2}$). Therefore the theoretical models can be classified/constrained by SSWGC through $\chi_{obs}$.  
\begin{equation}
 \frac{\xi^{4}_{V}}{\epsilon_{V}\,\eta^{2}_{V}} - \frac{\omega^{3}_{V}}{2\, \epsilon_{V}\,\eta_{V}} \geq \chi_{obs}
\end{equation}
\subsection{Dark energy scale fine tuning}
\noindent The initial conditions: Inflation ends when $\epsilon_{V}(a(x)_{\text{end}})=1$ or $\eta_{V}(a(x)_{\text{end}})\ll1$. Thus the solution of the $a(x)_{\text{end}}$ in equation(\ref{phisol}) would leads to initial dark energy density ($\rho(\phi_{i})$) at end of the inflation. 
\begin{equation}
 \frac{2 N_{e}}{(Nf_{a})^{2}}+\frac{1}{2} Ei\left[\left(\frac{a(x)_{\text{end}}}{Nf_{a}}\right)^{2}\right]-\ln[a(x)_{\text{end}}] = \frac{1}{2} Ei\left[\left(\frac{a(x)_{\text{ini}}}{Nf_{a}}\right)^{2}\right]-\ln[a(x)_{\text{ini}}] \label{phisol}
\end{equation}
 Here the inflaton field $a(x)_{end}$ represented as dark energy field $\phi_{i}$ at equipartition phase when $\Omega_{\phi_{i}}\approx10^{-4}$. Further we have arrived a condition $\phi_{i}< Nf_{a}/\sqrt{2}$ by minimizing the potential. Our case, the initial and present dark energy density can be calculated as follows,
 from table(\ref{tab:qpni}), we use $r=10^{-5}$ data which is $F_{a}=Nf_{a}=0.4\,M_{\text{Pl}}$ and corresponding field value $\phi_{i} = 0.195\,M_{\text{Pl}}$ with $M=10^{-2} \, M_{\text{Pl}}$, 
  $$ \rho(\phi_{i}) = V(\phi_{i}) = (1.0\times 10^{-2}\, M_{\text{Pl}})^{4} (1-\exp[-(0.195/0.4)^{2}]) = 7.02 \times 10^{64} \text{GeV}^{4}$$
  In the case of calculating present dark energy density, we can achieve $\rho(\phi_{0}) = (2.25\times 10^{-3} \text{eV})^{4}$ for $\Omega_{\phi_{0}} = 0.694$ by setting non-zero field value instead changing the symmetry breaking scale. The dynamically varying field would reduces the height of the potential which measures the dark energy scale at present as follows 
 \begin{equation}
 \rho(\phi_{0})  =  \rho_{c} \Omega_{\phi_{0}} = 3.67\times 10^{-47}\times 0.694 \times \text{GeV}^{4} =(2.25\times10^{-3} \text{eV})^{4}  \nonumber  
 \end{equation}
 The field value has to reach $\phi_{0} = 10^{-56}\, F_{a}$ as shown in the figure(\ref{fig:dde}) and we know that $F_{a} = 0.4 M_{\text{Pl}}$,
\begin{equation}
\rho(\phi_{0})  = V(\phi_{0})  =  (1.0\times 10^{-2}\, M_{\text{Pl}})^{4} (1-\exp[-(10^{-56}F_{a}/F_{a})^{2}]) = (2.4\times10^{-3} \text{eV})^{4} \nonumber 
\end{equation}
\noindent At this point we can conclude that $V(\phi_{0}) \approx \Lambda_{\text{DE}}^{4}$ achieved by ultra-light field value with ultra-light mass.  

%






\begin{thebibliography}{99}
\bibitem{guth} Alan H. Guth, \emph{Inflationary universe: A possible solution to the horizon and flatness problems}, {\PRD(23,1981,347)}.

\bibitem{linde} Andrei D. Linde, \emph{A New Inflationary Universe Scenario: A Possible Solution of the Horizon, Flatness, Homogeneity, Isotropy and Primordial Monopole Problems}, {\PL(B108,1982,389)}.

\bibitem{strucform} Pedro G. Ferreira, Michael Joyce, \emph{Structure Formation with a Self-Tuning Scalar Field}, {\PRL(79,1997,4740)}.

\bibitem{strucform1} Michael Doran, Georg Robbers, \emph{Early Dark Energy Cosmologies}, {\JCAP(06,2006,026)}.

\bibitem{strucform2} Michael Doran, Georg Robbers, Christof Wetterich, \emph{Impact of three years of data from the Wilkinson Microwave Anisotropy Probe on cosmological models with dynamical dark energy}, {\PRD(75,2007,023003)}.

\bibitem{strucform3} D. Huterer, D. Kirkby, R. Bean, A. Connolly, K. Dawson, et al., \emph{Growth of cosmic structure: probing dark energy beyond expansion}, {arXiv:1309.5385[astro-ph.CO]}.

\bibitem{planck1} Planck Collaboration, \emph{Planck 2018 results. X. Constraints on inflation}, {arXiv:1807.06211}.

\bibitem{planck2} Planck Collaboration, \emph{Planck 2018 results. I. Overview and the cosmological legacy of Planck}, {arXiv:1807.06205 [astro-ph.CO]}.

\bibitem{bao1} Wayne Hu, \emph{Dark Energy Probes in Light of the CMB}, {arXiv:astro-ph/0407158}.

\bibitem{bao2} P. A. R. Ade, et al., \emph{Planck 2015 results. XIV. Dark energy and modified gravity}, {arXiv:1502.01590 [astro-ph.CO]}.

\bibitem{ratra} B. Ratra, P. J. E. Peebles, \emph{Cosmological consequences of a rolling homogeneous scalar field}, {\PRD(37,1988,3406)}.

\bibitem{watteric} C.Wetterich , \emph{Cosmology and the fate of dilatation symmetry}, {\NP(B302,1988,668)}.

\bibitem{steinhardt1} R. R. Caldwell, R. Dave,  P. J. Steinhardt, \emph{Cosmological imprint of an energy component with general equation of state}, {\PRL(80,1998,1582)}.

\bibitem{dde1} Ivaylo Zlatev, Limin Wang, Paul J. Steinhardt, \emph{Quintessence, Cosmic Coincidence, and the Cosmological Constant}, {\PRL(82,1999,896)}.

\bibitem{watterich1} C.Wetterich , \emph{Inflation, quintessence, and the origin of mass}, {\NP(B897,2015,111)}.

\bibitem{soundspeed1} Simon DeDeo, R. R. Caldwell, Paul J. Steinhardt, \emph{Effects of the sound speed of quintessence on the microwave background and large scale structure}, {\PRD(67,2003,103509)}.

%
%
%
%

\bibitem{starobinsky} A. A. Starobinsky, \emph{A new type of isotropic cosmological models without singularity}, {\PL(B91,1980,99)}.

\bibitem{helical} Tianjun Li, Zhijin Li and Dimitri V. Nanopoulos, \emph{Helical phase inflation via non-geometric flux compactifications: from natural to Starobinsky-like inflation}, {\JHEP(10,2015,138)}.

\bibitem{ni} K. Freese, J. A. Frieman, A. V. Olinto, \emph{Natural Inflation with Pseudo Nambu-Goldstone Bosons}, {\PRL(65,1990,3233)}.

\bibitem{monodromy1} E. Silverstein, A. Westphal, \emph{Monodromy in the CMB: gravity waves and string inflation}, {\PRD(78,2008,106003)}.

\bibitem{monodromy2} Sergei Dubovsky, Albion Lawrence, Matthew M. Roberts, \emph{Axion monodromy in a model of holographic gluodynamics}, {\JHEP(02,2012,053)}.

\bibitem{hooft} G.'t Hooft, \emph{A planar diagram theory for strong interactions}, {\NP(B72,1974,461)}.

\bibitem{witten} Edward Witten, \emph{Theta dependence in the large N limit of four-dimensional gauge theories}, {\PRL(81,1998,2862)}.

\bibitem{pecci} R.D. Peccei, Helen R. Quinn, \emph{CP Conservation in the Presence of Instantons}, {\PRL(38,1977,1440)}.

\bibitem{pni1} Yasunori Nomura , Taizan Watari , Masahito Yamazaki,\emph{Pure natural inflation}, {\PL(B776,2018,227)}.

\bibitem{pni2} Yasunori Nomura, Masahito Yamazaki,\emph{Tensor modes in pure natural inflation}, {\PL(B780,2018,106)}.

\bibitem{pni3} Jeong-Pyong Hong, Masahiro Kawasaki, Masahito Yamazaki, \emph{Oscillons from pure natural inflation}, {\PRD(98,2018,043531)}.

\bibitem{carrol} S. Carrol, \emph{Quintessence and the rest of the world: suppressing long-range interactions}, {\PRL(81,1998,3067)}.


\bibitem{lyth1} David H. Lyth, \emph{What Would We Learn by Detecting a Gravitational Wave Signal in the Cosmic Microwave Background Anisotropy?}, {\PRL(78,1997,1861)}.

\bibitem{lyth2} George Efstathiou, Katherine J. Mack, \emph{The Lyth Bound Revisited}, {\JCAP(05,2005,008)}.

\bibitem{lyth3} Richard Easther, William H. Kinney, Brian A. Powell, \emph{The Lyth Bound and the End of Inflation}, {\JCAP(08,2006,004)}.

\bibitem{BBO1} E. S. Phinney et al., \emph{The Big Bang Observer}, {NASA Mission Concept Study (2003)}.

\bibitem{BBO2} G. M. Harry, P. Fritschel,D. A. Shaddock, W. Folkner, E. S. Phinney, \emph{Laser interferometry for the Big Bang Observer}, {Class. Quantum Grav. \textbf{23} (2006) 4887, [Erratum-ibid. \textbf{23} (2006) 7361]}.

\bibitem{BBO3} Latham A. Boyle, Paul J. Steinhardt, \emph{Probing the early universe with inflationary gravitational waves}, {\PRD(77,2008,063504)}.

\bibitem{BBO4} Sachiko Kuroyanagi, Christopher Gordon, Joseph Silk, and Naoshi Sugiyama, \emph{Forecast constraints on inflation from combined CMB and gravitational wave direct detection experiments}, {\PRD(81,2010,083524)}.  

\bibitem{nisymmetry1} Cristiano Germani , Alex Kehagias, \emph{UV-Protected Inflation}, {\PRL(106,2011,161302)}.

\bibitem{nisymmetry2} Djuna Croon, Veronica Sanz, \emph{Saving Natural Inflation}, {\JCAP(02,2015,008)}.

\bibitem{swampland1} Cumrun Vafa, \emph{The String Landscape and the Swampland}, {arXiv:hep-th/0509212}.

\bibitem{swampland2} Nima Arkani-Hamed, Lubos Motl, Alberto Nicolis, Cumrun Vafa, \emph{The String Landscape, Black Holes and Gravity as the Weakest Force}, {\JHEP(06,2007,060)}.

\bibitem{swampland3} Hirosi Ooguri, Cumrun Vafa, \emph{On the Geometry of the String Landscape and the Swampland}, {\NP(B766,2007,21)}.

\bibitem{swampland4} Hirosi Ooguri, Eran Palti, Gary Shiu, Cumrun Vafa, \emph{Distance and de Sitter Conjectures on the Swampland}, {\NP(B788,2019,180)}.
 
\bibitem{swampland5} Hitoshi Murayama, Masahito Yamazaki, Tsutomu T. Yanagida, \emph{Do We Live in the Swampland?}, {\JHEP(12,2018,032)}.

\bibitem{swampland6} Robert  Brandenberger, \emph{Initial conditions for Inflation - A short review}, {\IJMP(D26,2017,1740002)}.
 
\bibitem{tensfluc} N. Tsamis, R. P. Woodard, \emph{Relaxing the cosmological constant}, {\PL(B301,1993,351)}. 

\bibitem{scalarfluc1} R. H. Brandenberger, \emph{Back reaction of cosmological perturbations and the cosmological constant problem}, {hep-th/0210165}. 
    
\bibitem{scalarfluc2} G. Geshnizjani, R. Brandenberger, \emph{Back reaction of perturbations in two scalar field inflationary models}, {\JCAP(04,2005,006)}.

\bibitem{scalarfluc3} G. Marozzi, G. P. Vacca, R. H. Brandenberger, \emph{Cosmological back reaction for a test field observer in a chaotic inflationary model}, {\JCAP(02,2013,027)}.
      
\bibitem{irinstable1} A. M. Polyakov, \emph{Decay of Vacuum Energy}, {\NP(B834,2010,316)}.
 
\bibitem{irinstable2} D. Marolf, I. A. Morrison, \emph{The IR stability of de-sitter QFT: results at all orders}, {\PRD(84,2011,044040)}.       
      
\bibitem{ibanez} Luis E. Ibanez, Victor Martin-Lozano , Irene Valenzuela, \emph{Constraining neutrino masses,the cosmological constant and BSM physics from the weak gravity conjecture},{\JHEP(11,2017,066)}.

\bibitem{phenorsc} Hajime Fukuda, Ryo Saito, Satoshi Shirai, Masahito Yamazaki, \emph{Phenomenological consequences of the Refined Swampland Conjecture}, {\PRD(99,2019,083520)}. 

\bibitem{wgc1} Daniel Klaewer, Eran Palti, \emph{Super-Planckian spatial field variations and quantum gravity}, {\JHEP(01,2017,088)}.

\bibitem{wgc2} Eran Palti, \emph{The Weak Gravity Conjecture and Scalar Fields}, {\JHEP(08,2017,034)}.

\bibitem{wgc3} Dieter Lust, Eran Palti, \emph{Scalar Fields, Hierarchical UV/IR Mixing and The Weak Gravity Conjecture}, {\JHEP(02,2018,040)}.

\bibitem{sswgc} Eduardo Gonzalo, Luis E. Ibanez, \emph{A Strong Scalar Weak Gravity Conjecture and Some Implications}, {\JHEP(08,2019,118)}.

\bibitem{uvirphase1} Satoshi Shirai, Masahito Yamazaki, \emph{Is Gravity the Weakest Force?}, {arXiv:1904.10577 [hep-th]}.

\bibitem{uvirphase2} Alexander Kusenko, Volodymyr Takhistov, Masaki Yamada, Masahito Yamazaki, \emph{Fundamental Forces and Scalar Field Dynamics in the Early Universe}, {arXiv:1908.10930 [hep-th]}.

\bibitem{craig} Nathaniel Craig, \emph{Naturalness and New Approaches to the Hierarchy Problem}, {PiTP 2017 Lectures}.

\bibitem{selva} Selvaganapathy J., Partha Konar, Prasanta Kumar Das, \emph{Inferring the covariant $\Theta$-exact noncommutative coupling in the top quark pair production at linear colliders}, {\JHEP(06,2019,108)}.

\bibitem{seth} Nathaniel Craig, Seth Koren, \emph{IR Dynamics from UV Divergences: UV/IR Mixing, NCFT, and the Hierarchy Problem},{arXiv:1909.01365 [hep-ph]}.

\bibitem{instreheatdde1} G. N. Felder,  L. Kofman, A. D. Linde, \emph{Instant Preheating}, {\PRD(59,1999,123523)}.
 
\bibitem{instreheatdde2} A. H. Campos, H. C. Reis, R. Rosenfeld, \emph{ Preheating in Quintessential Inflation}, {\PL(B575,2003,151)}.

\bibitem{reheatdde1} Konstantinos Dimopoulos, Leonora Donaldson Wood, Charlotte Owen, \emph{Instant preheating in quintessential inflation with $\alpha$-attrators}, {\PRD(97,2018,063525)}

\bibitem{reheatdde2} Konstantinos Dimopoulos, Mindaugas Karciauskas, Charlotte Owen, \emph{Quintessential inflation with a trap and axionic dark matter}, {\PRD(100,2019,083530)}.

\bibitem{gravreheatdde1}  L. H. Ford, \emph{Gravitational particle creation and inflation}, {\PRD(35,1987,2955)}.

\bibitem{chiba1} Hiroyuki Tashiro, Takeshi Chiba, Misao Sasaki, \emph{Reheating after quintessential inflation and gravitational waves}, {Class. Quant. Grav. \textbf{21} (2004) 1761}

\bibitem{gravreheatdde2}  E. J. Chun, S. Scopel, I. Zaballa, \emph{Gravitational reheating in quintessential inflation}, {\JCAP(07,2009,022)}.

\bibitem{riccireheatdde1} K.  Dimopoulos, T.  Markkanen, \emph{Non-minimal gravitational reheating during kination}, {\JCAP(06,2018,021)}.
      
\bibitem{riccireheatdde2} T. Opferkuch, P. Schwaller, B. A. Stefanek, \emph{Ricci Reheating}, {\JCAP(07,2019,016)}.

\bibitem{dde2} G. Huey, L. Wang, R. Dave, R. R. Caldwell, P. J. Steinhardt, \emph{Resolving the cosmological missing energy problem}, {\PRD(59,1999,063005)}.

\bibitem{dde3} P. J. Steinhardt, L. Wang, I. Zlatev, \emph{Cosmological tracking solutions}, {\PRD(59,1999,123504)}.

\bibitem{dde4} P. J. Steinhardt, \emph{A quintessential introduction to dark energy}, {Phil. Trans. R. Soc. Lond. \textbf{A361} (2003) 2497}.

\bibitem{chiba} Takeshi Chiba, Antonio De Felice, Shinji Tsujikawa, \emph{Observational Constraints on Quintessence: Thawing, Tracker and Scaling models}, {\PRD(87,2013,083505)}.

\bibitem{saridakis} Md. Wali Hossain, R. Myrzakulov, M. Sami, Emmanuel N. Saridakis, \emph{Unification of inflation and dark energy $\grave{a}$ la quintessential inflation}, {\IJMP(D24,2015,1530014)}.

\bibitem{scherrer1}  Robert J. Scherrer, A. A. Sen, \emph{Thawing quintessence with a nearly flat potential}, {\PRD(77,2008,083515)}.

\bibitem{rrangarajan} Gaveshna Gupta, Raghavan Rangarajan, Anjan A. Sen, \emph{Thawing quintessence from the inflationary epoch to today}, {\PRD(92,2015,123003)}.

\bibitem{sugiy} Jean-Baptiste Durrive, Junpei Ooba, Kiyotomo Ichiki, Naoshi Sugiyama \emph{Updated observational constraints on quintessence dark energy models}, {\PRD(97,2018,043503)}

\bibitem{Hossain1} Md. Wali Hossain, R. Myrzakulov, M. Sami, Emmanuel N. Saridakis \emph{Variable gravity: A suitable framework for quintessential inflation},{\PRD(90,023512,2014)}.

\bibitem{Hossain2} Md. Wali Hossain, R. Myrzakulov, M. Sami, Emmanuel N. Saridakis \emph{Evading Lyth bound in models of quintessential inflation},{\PL(B737,2014,191)}.
  
\bibitem{sami} Chao-Qiang Geng, Chung-Chi Lee, R. Myrzakulov, M. Sami, Emmanuel N. Saridakis \emph{Observational constraints on varying neutrino-mass cosmology},{\JCAP(01,2016,049)}.

\bibitem{caldwell} R. R. Caldwell, E. V. Linder, \emph{Limits of Quintessence}, {\PRL(95,2005,141301)}.

\bibitem{scherrer2} Sourish Dutta, Robert J. Scherrer, \emph{Hilltop Quintessence}, {\PRD(78,2008,123525)}.
 
\bibitem{scherrer3} Sourish Dutta, Emmanuel N. Saridakis, Robert J. Scherrer, \emph{Dark energy from a quintessence (phantom) field rolling near potential minimum (maximum)}, {\PRD(79,2009,103005)}.

\bibitem{brax} Philippe Brax, Jerome Martin, \emph{Quintessence and Supergravity}, {\PL(B468,1999,40)}.
   
\bibitem{starobinskydde} Chao-Qiang Geng, Chung-Chi Lee, M. Sami, Emmanuel N. Saridakis, Alexei A. Starobinsky, \emph{Observational constraints on successful model of quintessential inflation}, {\JCAP(06,2017,011)}.   

\bibitem{dodelson} Kimberly Coble, Scott Dodelson, Joshua A. Frieman \emph{Dynamical $\Lambda$ models of structure formation}, {\PRD(55,1997,1851)}.

\bibitem{NNI} Kazuya Yonekura \emph{Notes on natural inflation}, {\JCAP(10,2014,054)}.

\end{thebibliography}
\end{document}